\documentclass{elsart}
\usepackage{epsfig}
\usepackage{amssymb}
\newcommand{\msun}{\mbox{$M_\odot$}}

\def\be{\begin{eqnarray}}
\def\ba{\begin{eqnarray}}
\def\ee{\end{eqnarray}}\def\ea{\end{eqnarray}}
\def\lsim{\mathrel{\rlap{\lower3pt\hbox{\hskip1pt$\sim$}}
     \raise1pt\hbox{$<$}}} %less than or approx. symbol
\def\gsim{\mathrel{\rlap{\lower3pt\hbox{\hskip1pt$\sim$}}
     \raise1pt\hbox{$>$}}} %greater than or approx. symbol
\def\la{\langle}\def\ra{\rangle}
\def\Tr{\rm Tr}
\def\del{\partial}
\def\cal{\it}
\def\bi{\bibitem}
\def\no{\nonumber\\}
\def\calL{\cal L}

\begin{document}

\runauthor{Brown, Lee, Rho}

\begin{frontmatter}
\title{Late Hadronization and Matter Formed at RHIC:\\ Vector Manifestation, BR Scaling and
Hadronic Freedom}

\author[suny]{Gerald E. Brown}
\author[suny] {Jeremy W. Holt}
\author[pnu,apctp]{Chang-Hwan Lee}
\author[saclay]{Mannque Rho}

\address[suny]{Department of Physics and Astronomy,\\
               State University of New York, Stony Brook, NY 11794, USA \\
(\small E-mail: gbrown@insti.physics.sunysb.edu}

\address[pnu]{Department of Physics, Pusan National University,
              Pusan 609-735, Korea\\
          (E-mail: clee@pusan.ac.kr)}
\address[apctp]{Asia Pacific Center for Theoretical Physics,
POSTECH, Pohang 790-784, Korea}

\address[saclay]{Service de Physique Th\'eorique,
 CEA Saclay, 91191 Gif-sur-Yvette
c\'edex, France (E-mail: mannque.rho@cea.fr)}

%\thanks[geb]{Ellen.Popenoe@sunysb.edu}
%\thanks[chl]{clee@pusan.ac.kr}
%\thanks[rho]{rho@spht.saclay.cea.fr}

\renewcommand{\thefootnote}{\fnsymbol{footnote}}
\setcounter{footnote}{0}

\begin{abstract}
Recent developments in our description of RHIC and related heavy-ion
phenomena in terms of hidden local symmetry theories are reviewed
with a focus on the novel nearly massless states in the vicinity of
-- both below and above -- the chiral restoration temperature $T_c$.
We present complementary and intuitive ways to understand both
Harada-Yamawaki's vector manifestation structure and Brown-Rho
scaling -- which are closely related -- in terms of ``melting" of
soft glues observed in lattice calculations and join the massless
modes that arise in the vector manifestation (in the chiral limit)
just below $T_c$ to tightly bound massless states above $T_c$. This
phenomenon may be interpreted in terms of the B\'eg-Shei theorem. It
is suggested that hidden local symmetry theories arise naturally in
holographic dual QCD from string theory, and a clear understanding of
what really happens near the critical point could come from a deeper
understanding of the dual bulk theory. Other matters discussed are
the relation between Brown-Rho scaling and Landau Fermi-liquid fixed
point parameters at the equilibrium density, its implications for
``low-mass dileptons" produced in heavy-ion collisions, the
reconstruction of vector mesons in peripheral collisions, the pion
velocity in the vicinity of the chiral transition point, kaon condensation viewed from the VM fixed point, nuclear physics with Brown-Rho scaling, and
the generic feature of dropping masses at the RGE fixed points in generalized
hidden local symmetry theories.
\end{abstract}

%\begin{keyword}
%\PACS{97.60.Lf; 97.80.Jp}
%\end{keyword}

\end{frontmatter}

\renewcommand{\thefootnote}{\arabic{footnote}}
\setcounter{footnote}{0}
%%%%%%%%%%%%%%%%%%%%%%%%%%%%%%%%%%%%%%%%%%%%%%%%%%%%%%%%%%%%%%%%%%%%%%%%%%%%%%%%

\section{Introduction\label{intro}}
The discovery at RHIC of what appears to be ``new matter" in the
form of a strongly interacting liquid, nothing like the quark gluon
plasma, brings up two issues: first, the long-standing ``old" issue of
what the structure of the state is in the vicinity of
the presumed chiral phase transition point as well as what the proper tool
to understand it is, and second, a ``new" issue as to what lies above the
critical point which the accepted theory of strong interactions,
QCD, is supposed to be able to access perturbatively. In this
review, we wish to address these issues in terms of an old idea on
in-medium hadron properties proposed in 1991 \cite{BR:91} which has
been recently rejuvenated with the surprisingly potent notion of
``vector manifestation (VM)" of hidden local symmetry
theory \cite{HY:PR} and buttressed with some recent results from
lattice QCD. Our principal thesis of this paper is that just as an
intricate and subtle mechanism is required to reach the VM structure
of chiral symmetry just below $T_c$ -- which is yet far from fully
understood -- from the standard linear sigma model picture
applicable (and largely established) at $T\sim 0$, the
structure of matter above and close to $T_c$ could also be intricate
and subtle from the starting point of QCD at $T\sim \infty$ at which
asymptotic freedom is applicable and ``established." We will make
here a leaping extension of the ideas developed in \cite{BGR06} in
which we infer from available information coming from lattice
results and hinted by RHIC data that the structures of matter just
below and just above the critical point can be related. The picture
we arrive at near the chiral restoration point, as remarked in
\cite{BR2004}, resuscitates the old ``B\'eg-Shei
theorem" \cite{beg-shei} which states: ``At short distances the
Nambu-Goldstone way merges with the Wigner-Weyl way: one can think
of symmetry without specifying the nature of the realization." The
picture we have developed is definitely falsifiable by lattice
calculations as well as by experiments. While awaiting the verdict,
we shall continue exploring the implications of this highly
attractive -- at least to us -- scenario.

As we will develop in this paper, the key to the possible new matter
produced at RHIC and its connection to the matter below $T_c$ is in
the glue from the gluons exchanged between quarks and its role in
chiral restoration as expressed in Brown-Rho scaling.
Our first task in this article is to rephrase the Harada-Yamawaki
theory  ``Hidden Local Symmetry at Loop" \cite{HY:PR} in terms of
some familiar results in chiral Lagrangian models in conjunction
with recent lattice results which we hope will be easier to
understand than the somewhat formal treatments given by Harada and
Yamawaki. Our description, being more intuitive, lacks rigor but
complements the gauge theoretic approach of Harada and Yamawaki. In
particular, we shall interpret their results ``pictorially" through
the Nambu-Jona-Lasinio (NJL) theory, which is an effective
theory possessing the symmetries of QCD, backed up
by the results of lattice gauge calculations carried out in full
QCD. We admit that there is sometimes ambiguity in how to interpret
lattice gauge simulation (LGS) through the NJL model, and so we will use
empirical data to make our interpretation believable. Of course, the
ultimate judge is how the Harada-Yamawaki theory describes
nature\footnote{In \cite{HY:PR}, only the $\rho$ meson enters as a
hidden gauge field. In two recent papers, Harada and
Sasaki~\cite{HS:a1} and Hidaka, Morimatsu and
Ohtani~\cite{Hidakaetal} independently showed that the axial vector
meson $a_1$ can also be suitably incorporated into the scheme. More
details will be given below.}. Their theory had many initial
surprises, such as ``hadronic freedom" (which we shall develop more
precisely later) as $T\rightarrow T_c$ from below, whereas the seemingly
equilibrated ratios of various hadrons emerging with temperature
$\sim T_c$ seemed to show that the interactions became stronger as
$T\rightarrow T_c$ from below.

One of the remarkable consequences of the notion of hadronic freedom
derived from the vector manifestation fixed point of HLS is that
certain processes can be more profitably described by fluctuating
around the VM fixed point with vanishing mass and coupling constants
instead of the standard practice of doing physics from the matter-free
vacuum. We will show what this implies in processes that exhibit chiral
symmetry in vacuum, the processes that take place near the chiral
transition point such as kaon condensation, pion velocity, etc.

%======================================================
\section{Soft Glue and the Vector Manifestation}

\subsection{Hard and Soft Glue}

Originally Brown-Rho scaling~\cite{BR:91} was proposed based on the
restoration of scale invariance as $T\rightarrow T_c$ or the density
$n\rightarrow n_c$ and was formulated in the skyrmion picture, which
models QCD in the large $N_c$ limit. This became, for a time,
somewhat complicated because clearly scale invariance is still
explicitly broken by the hard glue (which we often call epoxy). This
is just the gluon condensate found in the initial formulation of
Yang-Mills theory without quarks, the condensate which gives rise to
dimensional transmutation which produces the scale $\lambda_{QCD}$.
We now understand that the ``melting" of the soft glue is
responsible for the Brown-Rho scaling\footnote{This point was
implicit in the original formulation of Brown-Rho scaling in the
1991 paper but has remained unclear until recently. It was however
clarified in various publications in different terms (e.g.,
identifying the soft glue with quarkonium and the hard glue with
gluonium) before the lattice result~\cite{Miller00} came along. See,
e.g., the footnote 6 in ref.\ \cite{frs98}.}.

Even before Brown-Rho scaling was proposed in 1991, Su Houng
Lee~\cite{SHL} and Y. Deng \cite{Deng} found that about half of the
glue melted at $T_c(unquenched)$, the remainder (epoxy) remaining up
to beyond $T_c(quenched)$. The separate roles of the soft glue and
the epoxy which remained, the isolation of the black body radiation,
and other aspects were clarified in 1991 by Adami, Hatsuda, and Zahed
\cite{AHZ91} in their formulation of QCD sum rules at low
temperatures.

It became clear that the soft glue, which brings about a dynamical
breaking of scale invariance, was the agent building constituent
quarks out of the massless (in the chiral limit) current quarks and
holding them together in hadrons, whereas the hard glue, which
explicitly broke scale invariance, had nothing directly to do with the
hadronic masses. Thus, Brown-Rho scaling was accomplished by the
restoration of the dynamically broken scale invariance by the
melting of the soft glue. A theorem by Freund and Nambu \cite{FN68}
says that the dynamical breaking of scale invariance $requires$ an
explicit breaking. This shows the extreme subtlety in the interplay
of explicit breaking and spontaneous breaking of scale invariance --
the latter locked to quark condensates -- in contrast to other
global symmetries. Although no fully convincing proof exists, it is
however very reasonable that the dynamical breaking is restored as
$T\rightarrow T_c(unquenched)$ and the hadron masses go to zero.
However, the hard glue remains far above $T_c$.

%----------------------------------------------
We now describe how to understand the renormalization group results
of Harada and Yamawaki \cite{HY:PR} from the glue calculation in
LGS. Our point is simply that the glue which gives the quark its
mass, making it into a constituent quark, melts as $T\rightarrow
T_c$ from below, so that the constituent quark becomes a massless
current quark. Similarly, the mesonic exchange interactions between
hadrons result from the exchange of soft gluons, so that hadronic
interactions go to zero as $T$ goes up to $T_c$ from
below.
%\footnote{Question 1: The glue that ``melts," is it the
%condensate or fluctuating field? The lattice calculation measured
%the condensate $\la G^2\ra$ but the glue that is exchanged is the
%quantum (fluctuating) field of the soft glue (of the glue that
%breaks chiral symmetry dynamically). The vanishing of the condensate
%does not mean that the gluon exchange diagrams vanish. This point
%requires clarification.}
We will now walk the reader through these
arguments using results from lattice gauge simulations to clarify our
points.

\begin{figure}
\centerline{\epsfig{file=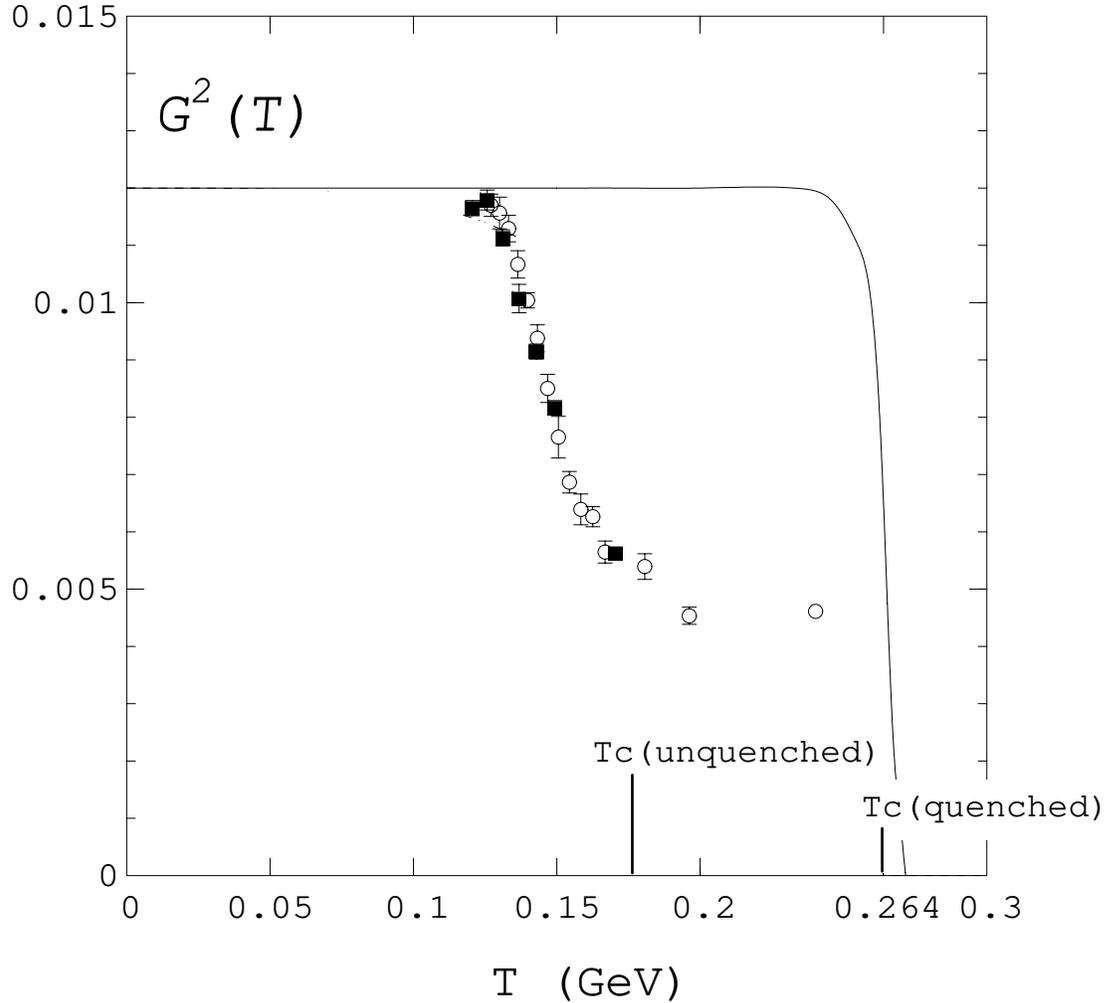,height=7.7in,
bbllx=85,bblly=240,bburx=484,bbury=611}}
\caption{Gluon
condensates taken from Miller~\cite{Miller00}. The lines show the
trace anomaly for SU(3) % (solid) and the ideal gluon gas (broken)
in comparison with that of the light dynamical quarks denoted by
the open circles and the heavier ones by filled circles.
Note that $G^2(T)$ ($=-\left(\beta(g)/2g^3\right) G_a^{\mu\nu} G_{\mu\nu}^a$)
is renormalization group invariant.
} \label{fig1}
\end{figure}

Dave Miller \cite{Miller00} put Fig. \ref{fig1} on the archives,
but was unable to publish it because the referee said that ``it
was well known." Yet it contains a great deal of important
information, which we now discuss. (Miller is preparing a Physics
Report on gluon condensates.)

%----------------------------------------------------------------
We carry out our discussion within the framework of the
Nambu-Jona-Lasinio (NJL) model as developed in \cite{BR2004}. One
can think of the NJL as arising when the light-quark vector mesons
(or more generally the tower of vector mesons as implied in
dimensionally deconstructed QCD~\cite{son-stephanov} or holographic
dual QCD that arises from string
theory~\cite{sakai-sugimoto,harvey}) and other heavy mesons are
integrated out. As such, it presumably inherits all the symmetries
of QCD as well as many of the results from hidden local symmetry theory,
which captures the essence of QCD~\cite{HY:PR}. It is well known that in
order to handle phase transitions in effective field theories, one
has to treat properly the quadratic divergences that are present in
loop graphs involving scalar fields. How this can be done in a
chirally invariant way is explained in \cite{HY:PR}. Now the cutoff
that occurs in the calculation represents the scale at which the
effective theory breaks down.
The natural scale for this is the chiral symmetry breaking scale
$\Lambda_{\chi SB} = 4\pi f_\pi\sim 1$ GeV. Brown and Rho
\cite{BR2004} suggested however that the cutoff in NJL should be at
$4\pi f_\pi/\sqrt 2 \sim 700$ MeV. This cutoff was thought to be
suitable for Wilsonian matching to constituent quarks rather than to
current quarks. For Wilsonian matching to QCD proper, the scale has
to be raised. In this regard, we are thinking in terms of a chiral
quark picture where constituent quarks and (pseudo)Goldstone bosons
coexist in a certain density or temperature regime. In addressing this
problem, Harada and Yamawaki \cite{HY:PR} were led to introduce
hidden local gauge invariance which allowed the vector meson mass to
be counted as of the same order in the chiral counting as the pion
mass, without however fermion degrees of freedom. We are proposing
that to be consistent with the Harada-Yamawaki theory, we need to
Wilsonian-match NJL to QCD at the same scale as in HLS theory. In
that case the loop graph with vector mesons comes in so as to cancel
the $\sqrt 2$ in the NJL cutoff denominator, so the Wilsonian
matching radius is raised to $4\pi f_\pi$.

NJL was carried out most neatly by Bernard, Meissner and
Zahed~\cite{BMZ:87}. We favor their results for a cutoff of $\Lambda=700$
MeV, which is close to $4\pi f_\pi/\sqrt 2$. In BGLR~\cite{BGLR} we got
our best fits\footnote{These parameters give $T_c=170$ MeV.} for
$\Lambda=660$ MeV and the NJL $G\Lambda^2=4.3$ (where $G$ is the
dimensionful coupling constant). In a mean-field type of mass
generation, it can be thought of as the coupling to constituent
quarks of the scalar\footnote{Not to be confused with the $\sigma$
that occurs in hidden local symmetry theory as the longitudinal
component of the $\rho$ meson.} $\sigma$-meson, $G\sim - g_{\sigma
QQ}^2/m_\sigma^2$.

At $n=0$, $T=0$, the proper variables are nucleons. We are sure of
this from the stunning success of nuclear structure theories. They
are bound states of three quarks, bound together by the glue. They
have mass $m_N$, mostly dynamically generated from the vacuum. The
degree of chiral symmetry breaking can be estimated by filling
negative energy states with the nucleons down to momentum scale
$\Lambda$. Thus
 \be {\rm B(glue)} = 4\int_0^\Lambda \frac{d^3 k}{(2\pi)^3}
\left\{\sqrt{k^2+m_N^2}- |\vec k| \right\}
 \ee
where we have subtracted  the perturbative energy $|\vec k|$. The
integral is easily carried out with the result
 \be {\rm B(glue)}
= 0.012\ {\rm GeV}^4,
 \ee
the value usually quoted for QCD sum rules. We used $\Lambda=660$
MeV.

As can be seen from Fig.~\ref{fig1}, there is no melting of the glue
until $T\approx 120$ MeV. The nucleon masses are just too heavy to
be pulled out of the negative energy sea by the thermal energies.
But as the temperature $T$ is increased, the nucleons will dissociate
into constituent quarks. Meyer, Schwenger and Pirner
\cite{Meyer2000} use a wave function which we write schematically
 \be
\Psi = Z | N\rangle + (1-Z^2)^{1/2} |3 q\rangle. \label{eq8} \ee In
other words there must be a transition of nucleons dissociating into
constituent quarks as mentioned. At this stage the glue which
surrounds the quarks starts to melt, and the curve for $G^2 (T)$
drops rapidly, down to $G^2\sim 0.0045$ at $T_c$. The heavy filled
circles are for bare quark masses which are 4 times greater than the
open MILC-collaboration ones, but the glue is insensitive to
explicit chiral symmetry breaking.

We would like to suggest that the lattice results imply in dense
medium a dissociation from nucleons to (colored) constituent (quasi)
quarks at some density above normal nuclear matter density but below
chiral restoration. This possibility was discussed a decade ago by
Alkofer, Hong and Zahed~\cite{alkoferetal} in terms of the NJL model
where the instability of a baryon skyrmion at a density higher than
normal was interpreted as splitting into $N_c$ (= 3 in nature)
constituent quarks, each with an effective mass dropping as a function of
density. It is not clear, though, that constituent quarks are stable
propagating degrees of freedom. The idea that the constituent quark
could be considered as a quark soliton (``qualiton") was proposed a
year earlier by Kaplan~\cite{kaplan-qualiton}, but it turned out that
no stable qualiton could be found from chiral Lagrangians so far
constructed~\cite{frishmanetal}. The question as to whether the
constituent quarks implied in the work of \cite{alkoferetal} are
bona-fide degrees of freedom near the chiral transition point as we
interpret on the basis of the lattice results remains unanswered. It
is interesting and intriguing to note that constituent quarks,
ill-defined in QCD language, seem to find a more precise definition
in holographic dual theory~\cite{myers}.

We can estimate the amount of soft glue that melts by changing
variables from nucleons to constituent quarks, where our degeneracy
factor is 12,
 \be {\rm B(soft\ glue)} = 12 \int_0^\Lambda
\frac{d^3k}{(2\pi)^3} \left(\sqrt{k^2 +m_Q^2}-|\vec k|\right). \ee
Taking $m_Q=320$ MeV, we find B(soft glue)$\sim \frac 12 (0.012)$
GeV$^4$. In the LGS the drop is a bit more than half of the $T=0$
glue. This could be achieved by choosing a slightly bigger
constituent quark mass, say, $m_Q\sim 370$ MeV.

So now at $T_c$ we are left with $G^2 (T) \sim 0.006$ GeV$^4$. This
is at $T_c$ where the soft glue has all melted and the constituent
quarks have become (massless) current quarks. Note that the next
point at $T\sim 1.4 T_c$ is equally high. There is no melting of the
glue between $T_c(unquenched)$ and $1.4 T_c(unquenched)$. This is
why we call this glue epoxy. It makes up the (colorless) Coulomb
interaction which binds the quark-antiquark molecules above $T_c$.
(Just above $T_c$ it binds them to zero or nearly zero mass. See
below.)

\begin{figure}[t]
\centerline{\epsfig{file=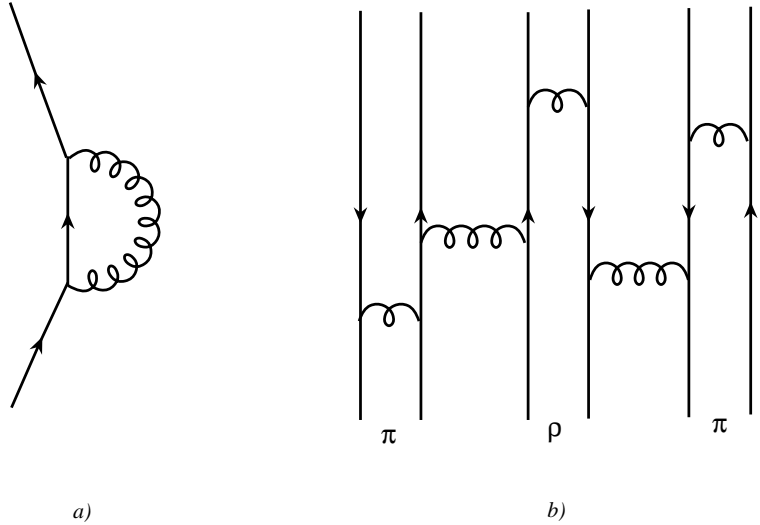,height=3.5in}}
%\vskip -20mm
%\centerline{\epsfig{file=fig4.eps,height=1.2in}}
\caption{Quark-gluon interactions. The wavy lines are gluons, the solid lines
are quarks. In a) the gluon helps build a constituent quark out of a current
quark. In b) part of the $2\pi \rightarrow \rho$ interaction is shown.
} \label{fig6ab}
\end{figure}

Pictorially, it is easy to see what happens. Assume that $T> 120$
MeV, where the nucleons have dissolved into constituent quarks. These
will have self energies, Fig.~\ref{fig6ab}a and interactions,
Fig.~\ref{fig6ab}b.
The constituent quark, whose mass can be thought of as the self
energy, will be converted to a massless current quark as the gluon
is melted. It is clear from this picture, and from the work of Harada
and Yamawaki \cite{HY:PR}, that the interaction between constituent
quarks, Fig.~\ref{fig6ab}b, will also go to zero as the soft glue is
melted.
%\footnote{Question2: Meaning the gauge coupling goes to zero?
%How does this happen in this picture? Here the exchanged gluons are
%quantum (fluctuating fields).}
The processes in Fig.~\ref{fig6ab}
result from fluctuating fields, not condensates. So they do not
follow directly from the melting of the condensate, but do follow
from the Harada-Yamawaki \cite{HY:PR} $G_V\rightarrow 0$ as
$T\rightarrow T_c$.

In general, both the gluon and the current quark will be colored,
also off-shell. Since they are only virtual particles they cannot
leave the fireball until $T=T_c$, with increasing energy. However,
they can interact with the pion, although the magnitude of the
interaction will be cut down, depending on how far off shell
they are. Upon reaching $T_c$ the color could, in principle,
escape. However, the system then goes into tightly bound colorless
chirally restored mesons. So, in the end, although they play a
role in equilibration, energy must be furnished, in the form of a
bag constant, to melt them.

The trace anomaly in the chiral limit is \be \theta_\mu^\mu =
-\frac{\beta(g)}{2g} G_a^{\mu\nu}(T) G^a_{\mu\nu}(T) \equiv G^2(T).
\ee In fact, the contribution from quarks comes as \be
\delta\theta^\mu_\mu =\sum_q {\bar m}_q \bar q q, \ee completely
from the explicit chiral symmetry breaking from the bare quark mass.
The bag constant $B$ is just
 \be B=\frac{1}{4} \theta_\mu^\mu.
\ee This shows how Brown-Rho scaling reflects the melting of the
soft glue. We shall return later to a discussion of how the LGS
support the scenario of meson masses going to zero as
$T\rightarrow T_c(unquenched)$. The Harada and Yamawaki work,
however, goes further and shows that the width of the $\rho$ to
$2\pi$ decay goes to zero as $T\rightarrow T_c$ from below:
 \be
\Gamma(\rho\rightarrow 2\pi)\propto \left(
\frac{\la\bar{q}q\ra_T}{\la\bar{q}q\ra_0}\right)^2\rightarrow 0
 \ee
just below $T_c$. With p-wave penetration factor, the power is 5
instead of 2, so it goes to zero faster. This was not foreseen by Brown
and Rho in \cite{BR:91} but is clearly true in our scenario of the
soft glue melting as $T\rightarrow T_c$ from below, as can be seen
from Fig.~\ref{fig6ab}b. %\footnote{Question 3: same as question 2.}
If the soft glue melts, then there is no transition from $\rho$ to
$2\pi$. In Section \ref{sec2} we will describe the chirally restored
mesons found above $T_c$, in the interval from $T_c$ to $\sim 2 T_c$
in quenched lattice gauge calculations, and later we will try to
connect these with the mesons such as $\pi$ and $\rho$ below $T_c$.

%-------------------------------------

\section{Generalized Hidden Local Symmetry (GHLS)}

The vector manifestation fixed point discovered by Harada and
Yamawaki~\cite{HY:PR} involved only the $\rho$ meson as a gauge field.
Now low-energy hadronic physics relevant to matter in medium
involves other ``heavy" hadrons such as the vector mesons $\omega$,
$a_1$ and the scalar $\sigma$. Indeed, in the discussion of the STAR
result for the $\rho^0/\pi^-$ ratio given below, we will invoke all
these mesons in $SU(4)$ multiplets. In confronting nature, it has
been assumed up until recently that the VM fixed point behavior, such as
the vanishing of the vector meson mass and the gauge coupling,
applies equally to these massive mesons as one approaches the
critical point. But how good is this assumption? An important part
of this question has recently been answered by Harada, and
Sasaki~\cite{HS:a1} and independently by Hidaka, Morimatsu and
Ohtani~\cite{Hidakaetal}, who studied what happens to the hidden
local symmetry structure of effective theories near the chiral phase
transition point when the $a_1$ degrees of freedom are considered
explicitly. This is an important issue in the modern development of
the field in two aspects. The first is that a generalized
hidden (flavor) local symmetry $naturally$ arises in holographic
dual QCD which emerges when the AdS/CFT conjecture is applied to
nonperturbative dynamics of QCD. The second is that the
$\rho$ and $a_1$ mesons make the (two) lowest members of the
infinite tower of massive gauge fields that descend via Kaluza-Klein
dimensional reduction in holographic dual QCD from a 5-D Yang-Mills
Lagrangian, and their role can be understood in a more general
context of QCD as a theory rather than as a particular modeling of
QCD. This suggests a strong theoretical backing of the multitude of
observations recently made by the authors in connection with RHIC
phenomena. In this section, we present a brief discussion of these
two developments.

\subsection{Infinite tower of vector mesons}\label{HDQCD}
In studying physical phenomena starting from the matter free vacuum (with
$T=n=0$), the local symmetry associated with the vector mesons that occur
in low-energy strong interaction physics is a
``luxury"~\cite{Georgi:2} that one can do without. In other words,
one does not have to have local symmetry when one deals with such
vector mesons as $\rho$, $\omega$ and $a_1$. The gauge symmetry is a
redundancy here. The symmetry is there since the physical field for
pions $U=e^{2i\pi/f_\pi}$ can be written in various different ways
introducing local fields. For instance, Harada and Yamawaki's HLS
theory uses the definition $U=\xi_L^\dagger\xi_R$ with
$\xi_{L,R}=e^{i\sigma (x)/f_\sigma}e^{\mp i\pi (x)/f_\pi}$, where the
$\sigma$ field is eaten by the gauge field which becomes massive.
There is a redundancy in that $U$ is invariant under the
multiplication by
 \be
\xi_{L,R}\rightarrow h(x)\xi_{L,R}.\label{h}
 \ee
This is quite analogous to what happens in other areas of physics.
For instance, in condensed matter systems in which one starts with
only electrons with short-ranged interactions, there
can be phases where the electron separates into a new fermion and a
boson~\cite{baskaran-anderson},
 \be
e(x)=b(x)f^\dagger (x).\label{electron}
 \ee
Under the local transformations,
 \be
b(x)&\rightarrow& e^{ih(x)} b(x),\nonumber\\
f(x)&\rightarrow& e^{ih(x)} f(x),
 \ee
the electron field remains unchanged. The new fields are redundant.
One can make the symmetry a gauge symmetry by introducing gauge
fields which are invisible in the original theory with the
electrons. This symmetry is an emerging one. Another example which
is quite analogous to this is the emergence of general coordinate
invariance in the AdS/CFT duality~\cite{horowitz-polchinski}.

The physics will be the same as that of a massive field with
$\sigma$ set equal to zero (corresponding to the unitary gauge) with
no gauge invariance. So why all these rigmaroles with local gauge
invariance and a hidden one at that? The reason is that there is a
power in doing physics with the gauge symmetry kept intact that is not
readily accessible in the gauge-fixed theory, and that involves
going up in scale with an effective low-energy theory.

How this works out is nicely described in \cite{georgi}. Imagine
doing a calculation approaching the scale corresponding to that of a
vector meson mass. Then having the local gauge invariance with the
Goldstone boson $\sigma$ in the Lagrangian facilitates two things.
First, with the Goldstone boson, one can locate where the EFT breaks
down, i.e., where ``new physics" shows up: It becomes strong
coupling at $\sim 4\pi m_V/g \sim 4\pi f$ where $f$ is the Goldstone
decay constant, $m_V$ is the vector boson mass and $g$ is the gauge
coupling. Without them, it is  complicated and awkward to locate the
break-down point. Now when an EFT breaks down, it is a signal that
the EFT is to be ``ultraviolet completed" to a fundamental theory,
which in our case is QCD. In Harada-Yamawaki theory this is
effectuated in some sense by the Wilsonian matching of correlators.

Second, one can systematically write higher-order terms as
powers of covariant derivatives,
 \be
\sim \frac{1}{16\pi^2}{\Tr}|D_\mu U|^4, \ \ \sim
\frac{1}{16\pi^2}{\Tr}|D^2 U|^2, \cdots
 \ee
In unitary gauge, these correspond to
 \be
\sim \frac{1}{16\pi^2}{\Tr} A^4, \ \ \sim
\frac{1}{16\pi^2}{\Tr}(\del A)^2, \cdots
 \ee
In the absence of symmetry guidance, it is difficult to write down
all the terms in the same power counting.

Although calculations may be more difficult, the above features can be
accounted for without gauge invariance for phenomenology at low
energy in the matter-free vacuum. This is the reason why one finds
the assertion in the literature that HLS, externally gauged massive
Yang-Mills and tensor field approaches are all equivalent. This
assertion is correct at the tree level~\cite{HY:PR}. However, the
situation is different when quantum loop effects are taken into
account. In particular, suppose one wants to do ``higher order"
calculations, say, in chiral perturbation theory. Then if the vector
meson mass needs to be considered as of the same scale as that of
the pion -- which is the case when Brown-Rho scaling is applicable
and the vector meson mass drops low as we believe in the high $T$ and/or
high density regime -- then a consistent chiral expansion involving
both vector mesons and pions is feasible {\it only when} local gauge
invariance is manifest. This point is the key point emphasized in
the work of Harada and Yamawaki.

The invariance (\ref{h}) involves one set of gauge fields, say,
$K=1$ vectors $\rho$ and $\omega$ with $h\in U(N_f)$. This way of
introducing gauge symmetry can be generalized to $K>1$ gauge fields.
The generalization is not unique and the different ways of
generalizing give rise to different theories. The simplest one is
the ``linear moose" structure on a lattice with a chain of ``link
fields" connecting the nearest-neighbor sites, i.e., gauge fields,
labelled by $K$ which can be extended to $\infty$. For instance, for
$K=2$, the two sites corresponding to $\rho$ and $a_1$ are connected
to each other by one link field and to the boundaries of L and R
chiral symmetries. Thus, there are towers of gauge fields for given
$K$'s. For a finite $K$, this ``moose" theory can be thought of as a
4-D gauge theory on a lattice with a finite lattice spacing, and in the
continuum limit with $K\rightarrow \infty$, it goes over to a 5-D YM
theory, with the extra dimension coming from the lattice. It has
been proposed that the resulting theory is a dimensionally
deconstructed theory of QCD~\cite{son-stephanov}. For a similar
discussion in a slightly different approach, see \cite{pomarol}.

The infinite tower of gauged vector mesons emerges also in
holographic dual QCD linked to the AdS/CFT duality in string
theory~\cite{sakai-sugimoto}. It has been shown that introducing
quark flavors (``probe branes") in the gravity sector in AdS space,
one finds a bulk theory that is thought to correspond to QCD in the
large 't Hooft limit ($\lambda=g\sqrt{N_C}\rightarrow \infty)$)
which comes out to be 5-D YM theory. When the fifth
dimension\footnote{The fifth dimension is called for by holography,
having to do with locality in energy in the renormalization group
equation. See ref.\ \cite{horowitz-polchinski} for a discussion on
this point.} is compactified with suitable boundary conditions, the
resulting Lagrangian is found to be a hidden local symmetry theory
(denoted in short as HDHLS for ``holographic dual hidden local
symmetry") with an infinite tower of vector mesons. This is an
effective theory valid below a Kaluza-Klein cutoff $M_{KK}$, the
compactification scale, possessing spontaneously broken chiral
symmetry and chiral anomalies (Wess-Zumino-Witten term) of QCD. In
going from 5-D to 4-D, the 5th component of the gauge field is
arbitrary. It turns out to be convenient to gauge fix it to zero to
make contact with Harada-Yamawaki HLS. However, if it is gauge-fixed
to the pion field, $all$ low-energy hadron processes, strong as well
as responses to the electroweak field, are found to be manifestly vector
dominated. Thus vector dominance (VD) is universal and automatic in
this theory. Surprisingly a variety of low-energy relations such as
KSRF, GMOR, etc.\ also come out correctly in this theory.

Another important aspect of the theory is that HDHLS, comprising
entirely of vector mesons and Goldstone bosons, has no fermion
degrees of freedom. This means that the ground-state baryons and
their excited states must arise as topological solitons, i.e.,
skyrmions, in 4-D theory or instantons in 5-D theory. It is
intriguing that the skyrmion is an indispensable ingredient in this
HDHLS theory. Therefore, in order to study dense matter where baryon
density must be taken into account, skyrmions have to be considered.
This point has been stressed in \cite{Parketal-skyrmion} but in
terms of the Skyrme Lagrangian, which we now know is not realistic
without incorporating vector mesons. It is likely that hidden local
gauge fields can provide topological order not present in the
non-gauged skyrmions~\cite{deconfinedQC}.

An interesting open problem is whether the $K=1$ theory of Harada
and Yamawaki, or the $K=2$ GHLS theory with $a_1$, can be understood
in terms of a truncated theory of a holographic dual HLS
theory. The latter predicts that the electromagnetic form factors of
the pion as well as the nucleon will be vector-dominated involving
the infinite tower of the vector mesons. In nature, it is known
empirically that the pion form factor is vector dominated by the
$K=1$ vectors but that the nucleon form factor is not dominated by
the lowest member
of vector mesons. It is not surprising that the vector
dominance involving an infinite tower of vector mesons could be
violated when the space is truncated to the lowest members, with the
violation representing the effect of the integrated-out vector
mesons. It is, however, intriguing that vector dominance with the
$K=1$ vector mesons holds so well for the pion form factors while it
does not for the nucleon form factors. It would be interesting to
investigate whether this rather special empirical observation follows
from HDHLS by integrating out the higher-lying members of the tower.
Such a study is in progress.

\subsection{How does the $a_1$ figure?}\label{GHLS}

For explaining some of the RHIC observations, e.g., the STAR
$\rho^0/\pi^-$ ratio discussed below, in addition to the pions and the
$K=1$ vector mesons ($\rho$ and $\omega$) other more
massive mesons in flavor $SU(4)$ symmetry need to be accounted for
near the critical region. As a first step to see how other degrees
of freedom enter in the fixed point structure of the EFT under
consideration, the role of $a_1$ has recently been elucidated
independently by two groups, Harada and Sasaki~\cite{HS:a1} and
Hidaka, Morimatsu and Ohtani~\cite{Hidakaetal}. It is found in
this ``generalized hidden local symmetry"
theory that as the order parameter of chiral symmetry, i.e., the
quark condensate $\la\bar{q}q\ra$, goes to zero (in the chiral
limit), there can be three different fixed points with $g=0$
characterized by
\begin{eqnarray}
&& \mbox{GL-type} \ : \
  M_\rho^2/M_{a_1}^2 \ \rightarrow\ 1 \,,
\nonumber\\
&& \mbox{VM-type} \ : \
  M_\rho^2/M_{a_1}^2 \ \rightarrow\ 0 \,,
\nonumber\\
&& \mbox{Hybrid-type} \ : \
  M_\rho^2/M_{a_1}^2 \ \rightarrow\ 1/3 \,.
\end{eqnarray}
Here the ``GL (Ginzburg-Landau)-type" corresponds to the standard
sigma model scenario, the ``VM-type" corresponds to the Harada-Yamawaki
vector manifestation scenario, and the ``hybrid-type" is a new
scenario that will be clarified below. These types are characterized
by different multiplet structures. To specify them, write the
representations of the scalar, pseudoscalar, longitudinal vector and
axial vector mesons as
\begin{eqnarray}
 |s \rangle
   &=& |(N_f,N_f^\ast) \oplus (N_f^\ast,N_f) \rangle\,,
\nonumber\\
 |\pi \rangle
   &=& |(N_f,N_f^\ast) \oplus (N_f^\ast,N_f) \rangle \sin\psi
\nonumber\\
&&{}+
       |(1,N_f^2-1) \oplus (N_f^2-1,1) \rangle \cos\psi\,,
\nonumber\\
 |\rho \rangle
   &=& |(1,N_f^2-1) \oplus (N_f^2-1,1) \rangle\,,
\nonumber\\
 |a_1 \rangle
   &=& |(N_f,N_f^\ast) \oplus (N_f^\ast,N_f) \rangle \cos\psi
\nonumber\\
&&{}-
       |(1,N_f^2-1) \oplus (N_f^2-1,1) \rangle \sin\psi\,,
\label{rep-mixing}
\end{eqnarray}
where $\psi$ denotes the mixing angle. Including the representation,
the fixed points are characterized by
\begin{eqnarray}
&& \mbox{GL-type} \ : \
  \cos \psi \ \rightarrow\ 0 \,,
\nonumber\\
&& \mbox{VM-type} \ : \
  \sin \psi \ \rightarrow\ 0 \,,
\nonumber\\
&& \mbox{Hybrid-type} \ : \
\nonumber\\
&& \qquad
  \sin \psi \ \rightarrow\ \sqrt{\frac{1}{3}} \,,
  \quad
  \cos \psi \ \rightarrow\ \sqrt{\frac{2}{3}} \,.
\end{eqnarray}
It comes out that these different fixed points predict different
couplings to the electromagnetic field. This is highly relevant for
phenomenological tests, e.g., in dilepton productions discussed
below. For instance, near the chiral phase transition point
the $\gamma\pi\pi$ coupling is predicted to approach
\begin{eqnarray} && \mbox{GL-type} \ : \
  g_{\gamma\pi\pi} \ \rightarrow\ 0 \,,
\nonumber\\
&& \mbox{VM-type} \ : \
  g_{\gamma\pi\pi} \ \rightarrow\ \frac{1}{2} \,,
\nonumber\\
&& \mbox{Hybrid-type} \ : \
  g_{\gamma\pi\pi} \ \rightarrow\ \frac{1}{3} \,.
\end{eqnarray}
Note that in the GL-type, VD gets restored as chiral symmetry is
restored, in stark contrast to the VM-type case for which VD is
strongly violated. The hybrid type also gives about $33\%$ violation
of the VD since the direct $\gamma\pi\pi$ comes to be $1/3$.

Now there are three fixed points, all consistent with chiral
restoration given by the RGE flow of the generalized hidden local
symmetry. The natural question is which fixed point is chosen by
nature when the system is driven to chiral restoration. In order to
answer this question, one would have to study the theory at high
temperature, density and with an increasing number of flavors. Work
is in progress on this matter. But a priori, there is nothing to
indicate that only one of them will be reached in nature by all
three conditions, high temperature, high density and high number of
flavors. Indeed it appears that the states above $T_c$, $n_c$ and
$N_f^c$ are of basically different nature, and furthermore, if other
soft modes such as kaon condensation appear before the chiral
restoration, the transition to the ``chirally restored" state will
be drastically modified from what is described by an effective
theory of HLS-type~\cite{deconfinedQC}.

%----------------------------------------------------------------

\section{Chirally Restored Mesons, Equivalently $\bar q q$ Bound States,
{} From $T_c$ To $2 T_c$} \label{sec2}

The question we wish to address next is why and how the mesons with
chiral symmetry spontaneously broken below $T_c$ reappear above
$T_c$. We shall see below that the $\pi, \sigma, \rho,$ and $a_1$, which
form a badly broken SU(4) below $T_c$,\footnote{Here $SU(4)$ is
``spin$\otimes$isospin" symmetry. In this
spin-isospin-symmetry-restored phase, we use the extended notation
$\pi \equiv (\pi,\eta)$ (pseudoscalar particles), $\sigma \equiv
(\sigma, \delta=a_0)$ (scalar particles), $\rho \equiv
(\rho,\omega)$ (vector particles), $a_1 \equiv (a_1, \epsilon=f_1)$
(axial-vector particles). Therefore, the number of meson states is
$2(chiral\ symmetry)\times 4(spin) \times 4(isospin) =32$ with
massive vector particles. For the massless case, we have only two
polarization states of vector particles, so the number of states is
$32\times 3/4=24$. In our estimates, however, due to the
fine-splitting which is caused by the spin-spin interaction, the vector
particles maintain a finite but small mass. Now, the
question is what happens when we approach the chiral restoration
temperature from below, where the vector-meson mass vanishes. We do
not have a clear answer for this question. In our approach, we assume
that the vector meson mass approaches zero but remains finite
until the chiral phase transition, at which point our mesonic bound states
take over.}
are nearly degenerate in the
region between $T_c (unquenched)$ and $T_c (quenched)$, a region
of 175 MeV $< T <$ 250 MeV. Brown et al. \cite{BJBP} argue that
mesons, rather than liberated quarks and gluons, are the correct
variables up to beyond $T_c (quenched)$ and that above $T_c
(unquenched)=T_{\chi SB}$ the (hard) glue remains condensed. Thus,
although mesons can be regarded as quark-antiquark pairs, each
quark must be connected with the antiquark in the meson by a
``string" (i.e., a line integral of the vector potential;
equivalently, a Wilson line) in order to preserve gauge
invariance. It was then noted that it is difficult to include
consequences of the line integral in the thermodynamic development
of the system.
Indeed, the Bielefeld LGS \cite{KKZP} have found that for temperatures $\gsim T_c$,
the confining properties of the heavy quark potential are just those of the
$T=0$ charmonium potential including string tension.
Now, in fact our very small mesons, with rms radius $\sim 0.2$ fm, are bound just above $T_c$ by the Coulomb plus magnetic interactions, but were
they not, the string tension would act as a backup to keep them small,
inside of the $\sim 0.5$ fm screening radius. In the above sense, all that
happens with chiral restoration is that the masses of the mesons go to zero.

We shall consider throughout only the mesons calculated in
Nambu-Jona-Lasinio, i.e., $\pi, \sigma, \rho$ and $a_1$ of the SU(4)
multiplet. These are collective excitations, analogous to nuclear
vibrations, in that their wave functions have many coherent
components; i.e., the wave functions involve sums over particle-hole
excitations, with momenta limited by a cutoff $\Lambda$, where
$\Lambda$ can be viewed as the Wilsonian matching scale for
constituent quarks. This is why these mesons tend to have strong
couplings. We believe that they are responsible for nearly all of
the thermodynamics, although there are many interesting effects such
as strangeness equilibration, etc., with which we do not deal.

The RHIC material should be ideally suited for descriptions in LGS.
There are nearly as many antiparticles as particles, so the baryon number is small. (It is zero in LGS.) \ Equilibration has been shown to be good at RHIC, at least down to $T_c$. The lattice calculations of the Coulomb potential are for heavy quarks, actually quarks of infinite mass. We are interested in
light-quark systems, $\pi, \sigma, \rho, a_1$, for which we have to
add magnetic effects. As pointed out in BLRS \cite{BLRS}, this can
be done by going back to the work of Brown \cite{Brown52}, who
showed that for stationary states of two $K$-electrons in heavy
atoms the interaction was of the form
 \be V_c =\frac{\alpha}{r}
\left(1-\vec\alpha_1\cdot\vec\alpha_2\right),
\label{eqn3}
 \ee
where the $\vec{\alpha}$'s are the Dirac velocity operators. Since
helicity is good above $T_c$, as nicely explained by
Weldon~\cite{weldon}, $\vec\alpha_1\cdot\vec\alpha_2=\pm 1$. Now
applying this to QCD, with opposite helicities for quark and antiquark,
we will have
 \be
V_c=\frac{2\alpha_s}{r}
 \ee
and for the same helicity $V_c=0$. The latter can then be neglected
for $T\sim T_c$.

\begin{figure}
\centerline{\epsfig{file=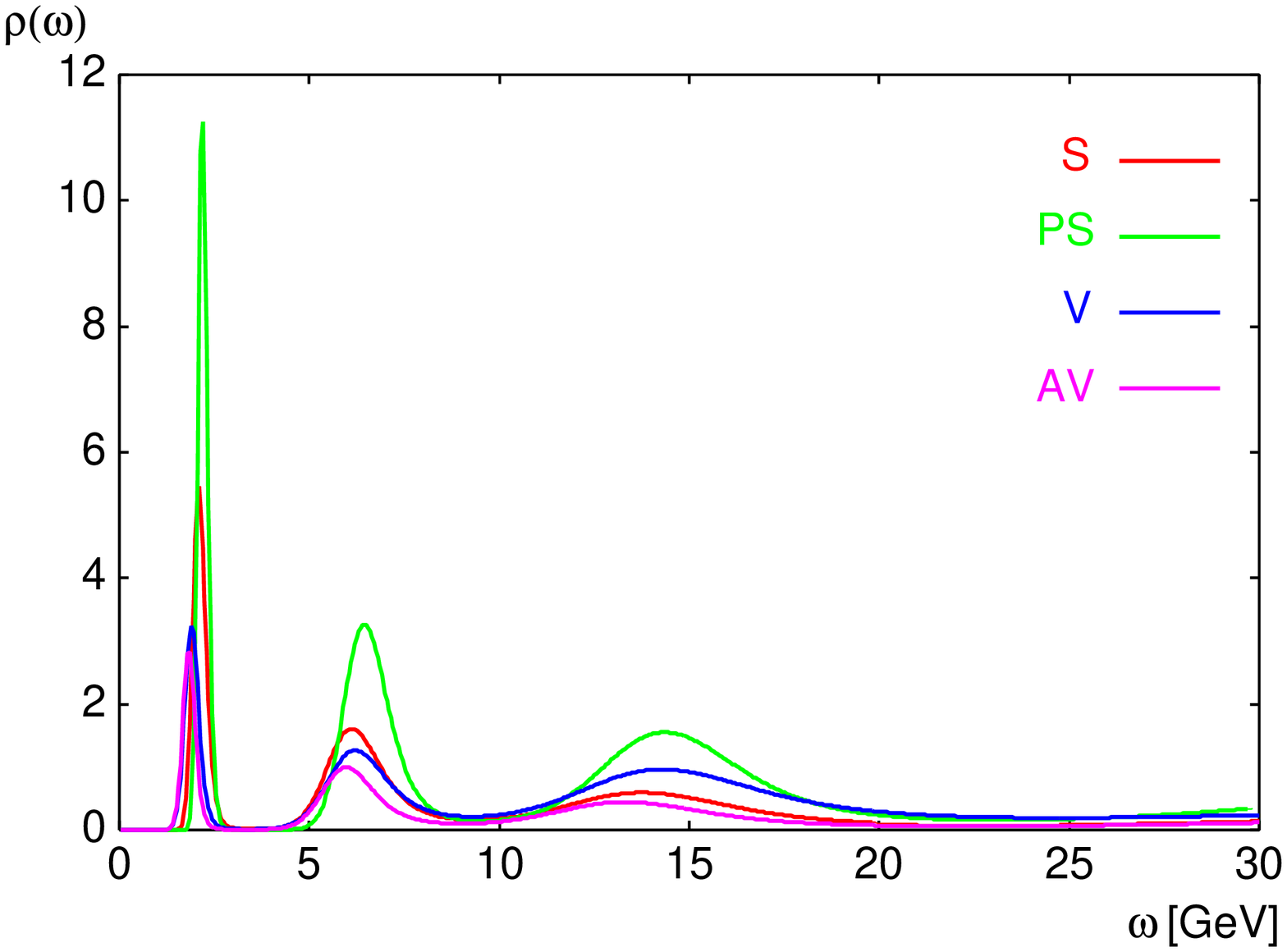,height=2in}
\epsfig{file=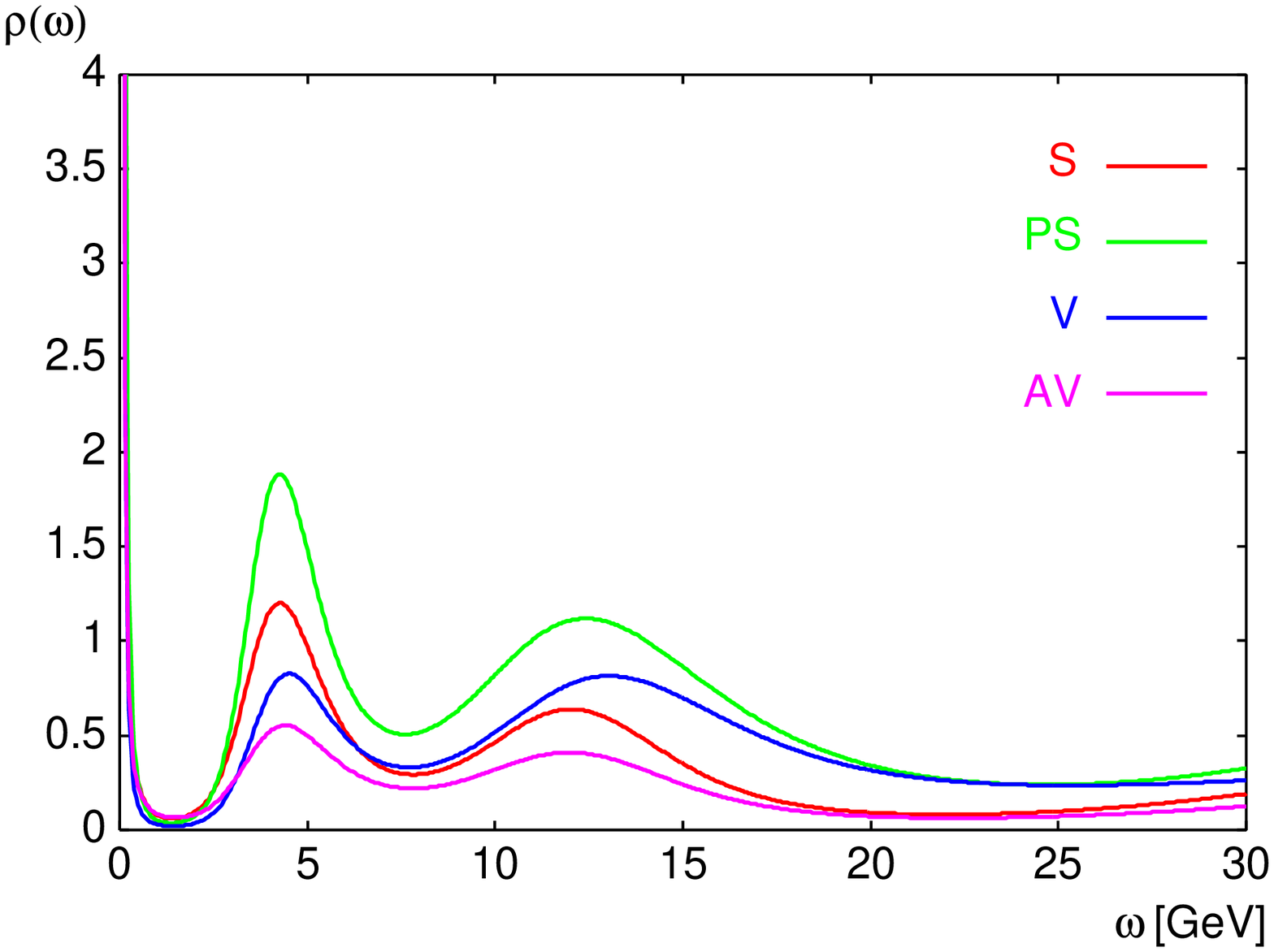,height=2in}}
\caption{Spectral functions of Asakawa et al.~\cite{asakawa03}.
Left panel: for $N_{\tau}=54$ ($T\simeq 1.4 T_c$).
Right panel: for $N_{\tau}=40$ ($T\simeq 1.9 T_c$).}
\label{figr1}
\end{figure}

Shuryak and Zahed \cite{SZ2} considered the region of $T\sim 2 T_c$,
where the $\bar q q$ bound states break up. We show in
Fig.~\ref{figr1} the lattice gauge calculations of the spectral
functions of Asakawa et al. \cite{asakawa03} for $T=1.4 T_c$ and for
$1.9 T_c$. We consider only the lowest excitations in each case.
Note that all SU(4) excitations are essentially degenerate. However,
the $T=1.9 T_c$ excitations at $\omega \sim 4.5$ GeV are rather
broad and it is clearly around here, roughly at the temperature
formed initially at RHIC, that the $\bar q q$ pairs are breaking up.
Now as they go through zero binding the $\bar q q$ scattering
amplitude goes from $\infty$ to $-\infty$ and the interaction
becomes very strong, as suggested by Shuryak and Zahed \cite{SZ2}.
In their formation the quark and antiquark velocities go to zero at
breakup, implying that our result Eq.~(\ref{eqn3}) holds only for
$T_c$ where the quark velocity is 1, the $\pi$ and $\sigma$ being
massless.
%\footnote{Question 5: I don't understand this point. It
%must be wrong. How can the helicity be a good quantum number so that
%(24) has eigenvalue $\pm 1$ when the quark (antiquark) velocity is
%zero?}
The result is that the viscosity is very low and this is the
material which is called ``perfect liquid"~\cite{son-viscosity}. We
do not have more to say other than that this is the upper end, in
temperature, of our $\pi, \sigma, \rho, a_1$ set of mesons. They
come unbound into quarks and antiquarks here.
Of course the colorless states we work with are mixed into a
multitude of other states, colorless and colored, by $T=1.9 T_c$.
This situation is quite analogous to the neutron giant resonances in
low-energy nuclear physics \cite{BD57}.

One might conclude from the quenched lattice results shown in
Fig.~\ref{figr1} that all 32 excitations labeled, S, PS, V and A
are degenerate. In fact, this is not entirely true. Chiral
symmetry restoration above $T_c$ does guarantee that the $\pi$ (in
the chiral limit) and $\sigma$ mesons are degenerate. However, the
$\rho$-meson will be slightly higher in energy than the $\pi$ and
$\sigma$, as will be the $a_1$, which is equivalent to the $\rho$
above $T_c$. The $\rho$ and $a_1$ have spin 1, the spin
interaction lifting their degeneracy with the $\pi$ and $\sigma$.
However, their magnetic moments \be \mu_{q, \bar q} =\mp
\frac{\sqrt{\alpha_s}}{p_0} \label{eq-mu} \ee are strongly
suppressed by the large magnitude of the thermal mass which enters
into $p_0$ \be p_0 =E+ m_{\rm th} \ee in the region of large
thermal masses just above $T_c$. In fact, we estimate that the
$\rho$ and $a_1$ will be shifted above the $\pi$ and $\sigma$ by
\be \Delta E=m_{\rm th}/6 \ee just above $T_c$.

%----------------------------------------------------------------

\section{Lattice Gauge Calculations in Full QCD}

LGS have been carried out in full QCD for $SU(2)\times SU(2)$ and are
published in two parts by O. Kaczmarek and F. Zantow~\cite{KZ1,KZ2}.
The first part \cite{KZ1} concerns a discussion of the quark-antiquark
free energies and zero temperature potential in two-flavor QCD.
The second part \cite{KZ2} concerns a detailed discussion of the lattice
data for the color singlet quark-antiquark internal energies.

Aside from rescaling the temperature from the earlier Bielefeld
quenched calculations to unquenched, the results are not very
different from the earlier calculations. This is perhaps not
surprising since the color singlet (Coulomb) interaction
dominates the phenomena; i.e., the color singlet gluon mode runs
the show.

Park, Lee and Brown \cite{PLB2005} showed that at $T_c$ putting the
earlier quenched Bielefeld LGS results into a Klein-Gordon equation
and doubling it in order to take into account the magnetic (Ampere's
Law) interaction as shown in the $\vec\alpha_1\cdot\vec\alpha_2$
term in Eq.~(\ref{eqn3}), the masses of the 32 degrees of freedom
shown in their Fig.~4 went to zero as $T$ went down to $T_c$. They
did not, however, include the spin-dependent interaction that
involves the $\mu_{q,\bar q}$ of Eq.~(\ref{eq-mu}). It is unclear
what takes place exactly at $T_c$;  indeed Harada and Yamawaki warned
against sitting on the fixed point~\cite{HY:PR}. However, the
thermal mass $m_{\rm th}$ is very large ($>$~1~GeV) when calculated
in the quenched approximation by Petreczky et al.\ \cite{Petreczky02}
at $T=1.5T_c$, not far above $T_c \sim 170$ MeV, and one expects the
quark ``mass" (more precisely the fourth component of the four
momentum) to go to $\infty$ with confinement. Thus a reasonable
assumption is that as $T$ goes down to $T_c$ from above, the
spin-dependent effects go to zero at $T_c$, because they are
inversely proportional to the ``mass'' (containing the large thermal mass),
and the apparent SU(4) symmetry
seen in the LGS above $T_c$ goes, just at $T_c$, over into an exact
SU(4). We give in the next section arguments from experiment that
this is true to the accuracy with which experiment can measure it,
and we show that this can explain what has been a surprising result up
to now.

\section{The STAR $\rho^0/\pi^-$ ratio}
As an illustration of the
potency of the notion of ``hadronic freedom," we sketch the
surprisingly simple argument that the STAR $\rho_0/\pi^-$ ratio in
peripheral $Au+Au$ collisions at RHIC is explained if the
$\pi,\sigma,\rho, a_1$ chirally restored mesons seen in LGS above
$T_c$ persist down through $T_c$, where they are essentially massless,
remain dormant until the temperature is low enough for them to go
back $\sim 90\%$ on shell, and then decay into pions. This is the
extension of the $\rho\rightarrow 2\pi$ decay described by Shuryak
and Brown \cite{SB:STAR}, in which it was shown that the
$\rho$-meson decayed into two pions at $T\sim 120$ MeV, thermal
freezeout for the peripheral experiments. The $\rho^0$ could be
reconstructed from the two pions, and it was shown that the $\rho$ mass had
decreased $\sim 10\%$, some of the decrease from Boltzmann factors,
but about 38 MeV from Brown-Rho scaling; i.e., as a medium effect
coming from the scalar densities furnished by the baryons and by the
vector mesons. The $\rho$-meson at the freezeout temperature was
only 10\% off-shell.

The result of STAR, after reconstructing the $\rho$-mesons by
following pion pairs back to their origin in the time projection
chamber, was that at $T\sim 120$ MeV the $\rho^0/\pi^-$ ratio was
\be \left.\frac{\rho^0}{\pi^-}\right|_{\rm STAR} =0.169 \pm 0.003
({\rm stat}) \pm 0.037 ({\rm syst}), \ee almost as large as the
$\rho^0/\pi^-=0.183 \pm 0.001({\rm stat})\pm 0.027({\rm syst})$ in
proton-proton scattering. The near equality of these ratios was not
expected\footnote{One might argue that the near equality means just
that nothing unusual with respect to pp scattering happens in the
STAR process, the ratio in heavy-ion collisions being about the same
as in pp scattering. Our description however involves the very
subtle notion of ``hadronic freedom" based on the vector
manifestation. This near equality could very well be coincidental
and could not be taken as an evidence against our scenario,
particularly since the ratio in pp scattering has not yet been
explained by particle theorists.}, since the $\rho$ meson width of
$\Gamma\sim 150$ MeV in free space is the strongest meson
rescattering that there is. If one assumes equilibrium at freezeout,
then the ratio is expected to be \cite{BMRS} \be \rho^0/\pi^- \sim
4\times 10^{-4}. \ee

We now show that the (nearly) massless $\pi, \sigma, \rho, a_1$ at
$T_c$ remain dormant until the temperature drops to $T\sim 120$ MeV,
which is the {\it freezeout temperature} for the peripheral
collisions. There the temperature, which we call the {\it flash
temperature} $T_{\rm flash}$, is such that the mesons go sufficiently
close to being on shell and their vector coupling approaches sufficiently close to the free space strength that they can decay into pions. Given our
$\pi, \sigma, \rho, a_1$ the $\rho^0/\pi^-$ ratio comes out close to
the empirical value. This shows that just below $T_c$ one has what
we call ``hadronic freedom," the $\pi, \sigma, \rho, a_1$ remaining
dormant. Then, at the flash temperature $T_{\rm flash}$ the $\sigma,
\rho$ and $a_1$ decay into pions, and it is a simple question of
counting in order to obtain the STAR $\rho^0/\pi^-$ ratio.

We first work around $T_{\rm flash}$ in order to show how nicely our
considerations here fit in with the Shuryak and Brown analysis
\cite{SB:STAR}. These authors noted that in the movement downwards
to $T_{\rm freezeout} \approx 120$ MeV the width of the $\rho$-meson
did not seem to change, although there should be a kinetic effect.
The negative mass shift automatically reduces (kinematically) the
width, both because of the reduced phase space and also due to the
power of $p$ in the $p$-wave matrix element. This kinematic shift in
width should go as inverse third power in shift in mass.

\begin{table}
\caption{$\Gamma^\star$ as function of temperature. For the point at
120 MeV we have switched over to the Shuryak and Brown
\cite{SB:STAR} value for $\Gamma^\star/\Gamma|_{\rho\rightarrow
2\pi}$. } \label{tab-gamma}
\begin{center}
\begin{tabular}{ccc}
\hline
T  & $m_\rho^\star/m_\rho$ & $\Gamma^\star/\Gamma$ \\
\hline
175 MeV &  0 & 0 \\
164 MeV & 0.18 & 0 \\
153 MeV & 0.36 & 0.01 \\
142 MeV & 0.54 & 0.05 \\
131 MeV & 0.72 & 0.22 \\
120 MeV & 0.90 & 0.67 \\
\hline
\end{tabular}
\end{center}
\end{table}

Now with the Harada-Yamawaki VM effect -- i.e., the intrinsic
background dependence -- taken into account, the width must drop
even faster as\footnote{The collisional width which we are not
considering here should also drop since the $a_1\rho\pi$ coupling
constant is also proportional to the gauge coupling $g_V$ which
drops.}
 \be
\frac{\Gamma_\rho^\star}{\Gamma_\rho}|_{\rho\rightarrow 2\pi} \sim
\left(\frac{m_\rho^\star}{m_\rho}\right)^3
\left(\frac{g_V^\star}{g_V}\right)^2 \Rightarrow
\left(\frac{m_\rho^\star}{m_\rho}\right)^5, \ee the dropping in
$(g_V^\star/g_V)$, from loop correction, beginning only a bit higher
up than $T_{\rm flash}=120$ MeV. If we use the scaling with third
power for $T=120$ MeV and that with fifth power above, we get the
results in Table~\ref{tab-gamma} for $\Gamma^\star$ as function of
temperature. We have let the mass drop linearly with temperature as
indicated by the more or less linear drop in Fig.~\ref{fig1} of the
soft glue with temperature. From Table~\ref{tab-gamma} we see that
at $T=T_{\rm flash}=120$ MeV the width $\Gamma^\star$ goes 2/3 of
the way back to the on-shell 150 MeV. The movement back towards the
on-shell value is rather sudden. This is why we call 120 MeV the
flash temperature. Thus, in the neighborhood of $T=120$ MeV the
mesons other than the pions, which are already present, decay into
pions.

Now the number of pions coming off at $T_{flash}$ is just a question of
counting. Once the meson is nearly on-shell it decays rapidly.
We find that in total 66 pions result at the end of the first
generation from the 32 $SU(4)$ multiplet, i.e., $\rho$ (18), $a_1$
(27), $a_0$ (4), $\pi$ (3), $\sigma$ (2) and $\epsilon\equiv f
(1285)$ (12), where the number in the parenthesis is the number of
pions emitted. Excluded from the counting are the $\omega$ and
$\eta$ since they leave the system before decaying. Leaving out the
three $\pi^-$'s coming from the $\rho^0$ decays which are
reconstructed in the measurement, we obtain
  \be
\frac{\rho^0}{\pi^-}\approx 3/(22-3)\approx 0.16.\label{prediction}
 \ee
We can understand this large ratio (very large compared with the
equilibrium $4\times 10^{-4}$) by the fact that the mesons go
through a hadron free region until they decay, never equilibrating.
With the same $\pi, \sigma, \rho, a_1$ that we see above $T_c$ in
the quenched LGS, we find the right number of pions are emitted to
fit the STAR experiment.

We should stress that whereas $T_{flash}$ is independent of
centrality, $T_{freeze\ out}$ is lower for higher centrality. Thus,
$\rho$'s cannot be reconstructed from the central collisions in
STAR. The reason is simple: the pions which they have decayed into
will have suffered rescatterings following the decay.

%---------------------------------------------------------------

\section{Comments on Hanbury Brown-Twiss Puzzle}
\label{sec9}

We have given a detailed description of the dynamics of what is
usually called the ``mixed phase," at least of the phase from $T_c$
down to $T_{\rm flash}$. For the peripheral collisions involved
here, the pions are emitted by the vector and axial vector mesons
and then leave the system at $T_{\rm freezeout}$ without interacting
with the fireball as a whole. It is hard to see how these will carry
information about the latter.

\begin{figure}
\centerline{\epsfig{file=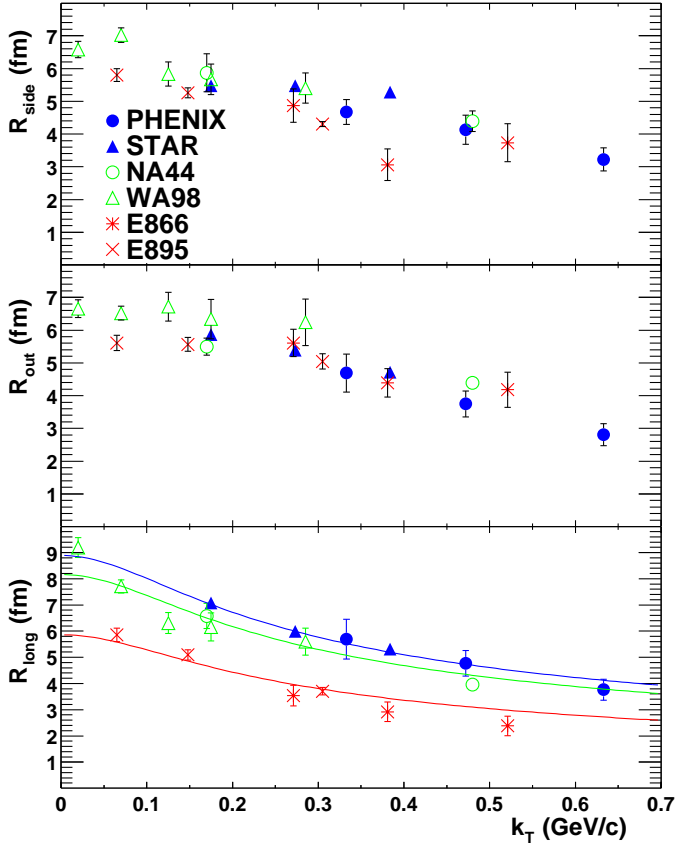,height=4.5in}}
\caption{The measurements of HBT radii for pion pairs
(taken from Fig.~2 of Adcox et al. \cite{PHENIX:HBT})
by PHENIX \cite{PHENIX:HBT}, STAR \cite{STAR:HBT},
NA44 \cite{NA44:HBT}, WA98 \cite{WA98:HBT}, E866 \cite{E866:HBT}
and E895 \cite{E895:HBT}.
These show all three radii to be essentially the same.
The bottom plot includes fits to the data.
The data are for $\pi^-$ results except for the NA44 results, which
are for $\pi^+$.
}
\label{FigXX}
\end{figure}

With respect to central collisions the situation is different, in
that the pions are emitted for temperatures greater than $T_{\rm
freezeout}$. We suggest that there may still be some influence on
the HBT puzzle because of how the dynamics are affected by the vector
and axial-vector mesons having to go on-shell before they can
interact (and decay). This means that in the entire mixed phase from
$T=T_c \simeq 175$ MeV down to $T \simeq 120$ MeV the expansion will be at nearly
the velocity of light $c$, which it begins with at $T_c$. This is
because due to hadronic freedom there is essentially no interaction
with the off-shell mesons until $T$ has dropped to $\sim 120$ MeV,
only with a few of the melted soft gluons. This means that the
system can increase in radius $\sim 5$ fm, one fermi for each
interval in Table~\ref{tab-gamma}. Thus, the system will not be very
far from spherical, at least pumpkin shaped, by the time the pions
begin interacting.

Once the vector and axial-vector mesons are on-shell at $T\sim
120$ MeV, there will be an explosion because their interactions are
suddenly turned on. Given an exploding, nearly spherical system of
large radius, most natural would seem to be outward, sideways and
longitudinal radii which are similar (See Fig.~\ref{FigXX}).
In any case we believe that the HBT calculations should be carried out
starting from our detailed dynamics.

\section{The Pion Velocity at $T_c$}
There are several quantities measured in laboratory experiments and
lattice simulations that can eventually be checked against. Here we
treat the pion velocity. Possible deviation from the velocity of
light for the massless pion (in the chiral limit) is expected since
Lorentz invariance is broken in a heat bath. The relevant Lagrangian
with Lorentz symmetry broken is
 \ba \tilde{\cal L} &=&\biggl[
  (F_{\pi,{\rm bare}}^t)^2 u_\mu u_\nu
  +
  F_{\pi,{\rm bare}}^t F_{\pi,{\rm bare}}^s
    \left( g_{\mu\nu} - u_\mu u_\nu \right)
\biggr] \mbox{tr} \left[
  \hat{\alpha}_\perp^\mu \hat{\alpha}_\perp^\nu
\right]\no &&+~\biggl[
 (F_{\sigma,{\rm bare}}^t)^2 u_\mu u_\nu
 + F_{\sigma,{\rm bare}}^t F_{\sigma,{\rm bare}}^s
    \left( g_{\mu\nu} - u_\mu u_\nu \right)
\biggr] \mbox{tr} \left[
  \hat{\alpha}_\parallel^\mu \hat{\alpha}_\parallel^\nu
\right]
\nonumber\\
&& +~ \Biggl[
  - \frac{1}{ g_{L,{\rm bare}}^2 } \, u_\mu u_\alpha g_{\nu\beta}
  - \frac{1}{ 2 g_{T,{\rm bare}}^2 }
  \left(
    g_{\mu\alpha} g_{\nu\beta}
   - 2 u_\mu u_\alpha g_{\nu\beta}
  \right)
\Biggr] \, \mbox{tr} \left[ V^{\mu\nu} V^{\alpha\beta}
\right]+\cdots \ ,
\nonumber\\
\label{Lag:no-L}
\end{eqnarray}
where $F_{\pi,{\rm bare}}^t$ ($F_{\sigma,{\rm bare}}^t$) and
$F_{\pi,{\rm bare}}^s$ ($F_{\sigma,{\rm bare}}^s$) denote the {\it
bare} parameters associated with the temporal and spatial decay
constants of the pion (of the $\sigma$). Here $u=(1,0)$ is the unit
four-vector for the rest frame. The parameters of the Lagrangian are
the ``bare" ones determined at the matching point by matching HLS
correlators to QCD ones. We recall here that, due to the Lorentz
symmetry violation, the two variables $\xi_{\rm L}$ and $\xi_{\rm R}$
included in the 1-forms $\hat{\alpha}_\perp^\mu$ and
$\hat{\alpha}_{\parallel}^\mu$ in Eq.~(\ref{Lag:no-L}) are
parameterized as
\begin{equation}
\xi_{\rm L,R} = e^{i\sigma/F_\sigma^t} e^{\mp i\pi/F_\pi^t} ,
\end{equation}
where $F_{\pi,{\rm bare}}^t$ and $F_{\sigma,{\rm bare}}^t$ are the
bare parameters associated with the temporal decay constants of the
pion and the $\sigma$.

We also need the terms of ${\cal O}(p^4)$ for the present analysis:
\begin{eqnarray}
\bar{\mathcal L}_{z_2} = \left[
  2 z_{2,{\rm bare}}^L u_\mu u_\nu g_{\nu\beta}
  + z_{2,{\rm bare}}^T
    \left(
      g_{\mu\alpha}g_{\nu\beta} - 2 u_\mu u_\alpha g_{\nu\beta}
    \right)
\right] \, \mbox{tr} \left[
    \hat{\mathcal A}^{\mu\nu} \hat{\mathcal A}^{\alpha\beta}
\right] \ , \label{Lag:no-L:z}
\end{eqnarray}
where the parameters $z_{2,{\rm bare}}^L$ and $z_{2,{\rm bare}}^T$
correspond in medium to the vacuum parameter $z_{2,{\rm
bare}}$~\cite{HY:PR} at $T=\mu=0$. $\hat{\mathcal A}^{\mu\nu}$ is
defined by
\begin{equation}
\hat{\mathcal A}^{\mu\nu} = \frac{1}{2} \left[
  \xi_{\rm R} {\mathcal R}^{\mu\nu} \xi_{\rm R}^{\dag}
  -
  \xi_{\rm L} {\mathcal L}^{\mu\nu} \xi_{\rm L}^{\dag}
\right] \ ,
\end{equation}
where ${\mathcal R}^{\mu\nu}$ and ${\mathcal L}^{\mu\nu}$ are the
field-strength tensors of the external gauge fields ${\mathcal
R}_\mu$ and ${\mathcal L}_\mu$:
\begin{eqnarray}
{\mathcal R}^{\mu\nu} &=&
  \partial^\mu {\mathcal R}^\nu - \partial^\nu {\mathcal R}^\mu
  - i \left[ {\mathcal R}^\mu \,,\, {\mathcal R}^\nu \right]
\ ,
\nonumber\\
{\mathcal L}^{\mu\nu} &=&
  \partial^\mu {\mathcal L}^\nu - \partial^\nu {\mathcal L}^\mu
  - i \left[ {\mathcal L}^\mu \,,\, {\mathcal L}^\nu \right]
\ .
\end{eqnarray}

Now define the parametric $\pi$ and $\sigma$ velocities as
\begin{equation}
 V_\pi^2 = {F_\pi^s}/{F_\pi^t}, \qquad
 V_\sigma^2 = {F_\sigma^s}/{F_\sigma^t}.
\end{equation}
The approach to the chiral restoration point should be characterized
by the equality between the axial-vector and vector current
correlators in QCD, $G_A - G_V \to 0$ for $T \to T_c$. The EFT
should satisfy this also for any values of $p_0$ and $\bar{p}$ near
the matching point provided the following conditions are met:
$(g_{L,{\rm bare}}, g_{T,{\rm bare}}, a_{\rm bare}^t, a_{\rm
bare}^s) \to (0,0,1,1)$ for $T \to T_c$. As in dense medium, this
implies that at the tree or bare level, the longitudinal mode of the
vector meson becomes the real NG boson and couples to the vector
current correlator, while the transverse mode decouples. A
non-renormalization theorem by Sasaki~\cite{NonRen-sasaki} shows
that $(g_L, a^t, a^s) = (0,1,1)$ is a fixed point of the RGEs
satisfied at any energy scale. Thus the VM condition is given by
\begin{eqnarray}
 (g_L, a^t, a^s) \to (0,1,1) \quad \mbox{for}\quad T \to T_c.
\label{evm}
\end{eqnarray}
The VM condition for $a^t$ and $a^s$ leads to the equality between
the $\pi$ and $\sigma$ (i.e., longitudinal vector meson) velocities:
\begin{eqnarray}
 \bigl( V_\pi / V_\sigma \bigr)^4
 = \bigl( F_\pi^s F_\sigma^t / F_\sigma^s F_\pi^t \bigr)^2
 = a^t / a^s
 \stackrel{T \to T_c}{\to} 1.
\label{vp=vs:2}
\end{eqnarray}
This is easy to understand in the VM scenario since the longitudinal
vector meson becomes the chiral partner of the pion. This equality
holds at $T_c$ whatever the value of the bare pion velocity obtained
at the matching point.

\vskip 0.2cm $\bullet$ {\it The standard sigma model scenario}

\vskip 0.2cm

Before we get into the discussion on the HLS prediction, it is
instructive to see what we can expect in chiral models without light
vector meson degrees of freedom. Here the basic assumption is that
near chiral restoration, there is no instability in the channel of
the degrees of freedom that have been integrated out. In this
pion-only case, the appropriate effective Lagrangian for the axial
correlators is the in-medium chiral Lagrangian dominated by the
current algebra terms,
 \be
{\calL}_{eff}=\frac{{f_\pi^t}^2}{4}\left({\Tr}\nabla_0 U\nabla_0
U^\dagger - v_\pi^2{\Tr}\del_i U\del_i U^\dagger\right) -\frac 12
\chi {\rm Re} M^\dagger U\label{LA}+\cdots\label{Leff}
 \ee
where $v_\pi$ is the pion velocity, $M$ is the mass matrix
introduced as an external field, $U$ is the chiral field and the
covariant derivative $\nabla_0 U$ is given by $\nabla_0 U=\del_0 U
-\frac i2 \mu_A (\tau_3 U +U\tau_3)$ with $\mu_A$ the axial isospin
chemical potential. The ellipsis stands for higher order terms in
spatial derivatives and covariant derivatives.

The quantities that we need to study are the vector isospin
susceptibility (VSUS) $\chi_V$ and the axial-vector isospin
susceptibility (ASUS) $\chi_A$ defined in terms of the vector charge
density $J^0_a (x)$ and the axial-vector charge density $J^0_{5a}
(x)$ by the Euclidean correlators:
 \be
\delta_{ab}\chi_V&=& \int^{1/T}_0 d\tau\int d^3\vec{x}\la J^0_a
(\tau, \vec{x}) J^0_b (0,\vec{0})\ra_\beta,\\
\delta_{ab}\chi_A&=& \int^{1/T}_0 d\tau\int d^3\vec{x}\la J_{5a}^0
(\tau, \vec{x}) J_{5b}^0 (0,\vec{0})\ra_\beta
 \ee
where $\la ~\ra_\beta$ denotes thermal average and
 \be
J_a^0\equiv \bar{\psi}\gamma^0\frac{\tau^a}{2}\psi, \ \
J_{5a}^0\equiv \bar{\psi}\gamma^0\gamma^5\frac{\tau^a}{2}\psi
 \ee
where $\psi$ is the quark field and $\tau^a$ is the Pauli matrix
generator of the flavor $SU(2)$. Given the effective action
described by (\ref{Leff}), with possible non-local terms ignored,
the axial susceptibility (ASUS) takes the simple form
 \be
\chi_A=-\frac{\del^2}{\del\mu_A^2}{\calL}_{eff}|_{\mu_A=0}={f_\pi^t}^2.
 \ee
The principal point to note here is that {\it as long as the
effective action is given by local terms (subsumed in the ellipsis)
involving the $U$ field, this is the whole story}: There is no
contribution to the ASUS other than the temporal component of the pion
decay constant.

Next one assumes that at the chiral phase transition point $T=T_c$,
the restoration of chiral symmetry dictates the equality
 \be
\chi_A=\chi_V.
 \ee
While there is no lattice information on $\chi_A$, $\chi_V$ has been
measured as a function of temperature~\cite{gottliebetal,BR:96}. In
particular, it is established that
 \be
\chi_V|_{T=T_c}\neq 0,
 \ee
which leads to the conclusion that
 \be
f_\pi^t|_{T=T_c}\neq 0.
 \ee
On the other hand, it is expected and verified by lattice
simulations that the space component of the pion decay constant
$f_\pi^s$ should vanish at $T=T_c$. One therefore arrives at
 \be
v_\pi^2\sim f_\pi^s/f_\pi^t\rightarrow 0, \ \ T\rightarrow
T_c.\label{sigmamodel}
 \ee
This is the main conclusion of the standard chiral
theory~\cite{ss-sus}.

What this means physically is as follows. The pole mass of the pion
$m_\pi^p$ in a heat bath is related to the screening mass $m_\pi^s$
via ${m_\pi^p}^2=v_\pi^2 ({m_\pi^s}^2 +\vec{k}^2)$. Thus the
vanishing of the pole mass would imply in this scenario the
vanishing of the pion velocity. In some sense this result would
indicate a maximal violation of Lorentz invariance and this will be
at a stark variance with what we find in HLS/VM theory described
below.

This elegant argument has a caveat. If one uses the same argument
for the VSUS, one gets a wrong answer. The effective Lagrangian for
calculating the vector correlators is of the same form as the ASUS,
Eq.~(\ref{Leff}), except that the covariant derivative is now
defined with the vector isospin chemical potential $\mu_V$ as
$\nabla_0 U=\del_0 U-\frac 12 \mu_V (\tau_3 U-U\tau_3)$. Now if one
assumes as done above for $\chi_A$ that possible non-local terms can
be dropped, then the VSUS is given  by
 \be
\chi_V=-\frac{\del^2}{\del\mu_V^2}{\calL}_{eff}|_{\mu_V=0}
 \ee
which can be easily evaluated from the Lagrangian. One finds that
 \be
\chi_V=0
 \ee
{\it for all temperatures.} While it is expected to be zero at $T=0$,
the vanishing of $\chi_V$ for $T\neq 0$ is at variance with the lattice
data at $T=T_c$.

The sigma model prediction (\ref{sigmamodel}) can be simply
understood from the fact that in the absence of other degrees of
freedom, $\chi_A$ is directly related to $f_\pi^t$ and $\chi_A$
is equal to $\chi_V$ at the chiral restoration point. Since $\chi_V$
is seen to be nonzero, $f_\pi^t$ does not vanish at $T_c$ whereas
the space component $f_\pi^s$ does. Now one can ask why $\chi_A$
should be given entirely by $f_\pi^t$ at $T_c$. There is no reason
why there should not be some additional contributions to $\chi_A$
other than from $f_\pi^t$. Indeed, this is the defect of the pion-only
sigma model scenario. We will see below that when the $\rho$ meson goes
massless at $T_c$, the longitudinal component of the $\rho$ meson
contributes to $\chi_A$ on the same footing as the $\pi$ and hence
the observation that the non-vanishing of $\chi_V$ implies
non-vanishing of $f_\pi^t$ is invalidated.

\vskip 0.2cm

$\bullet$ {\it The HLS/VM scenario} \cite{HKRS:pv}

\vskip 0.3cm

In the presence of the $\rho$ meson in HLS, it comes out that
$(f_\pi^t, f_\pi^s)\rightarrow (0, 0)$ and $\chi_A\rightarrow
\chi_V\neq 0$ as $T\rightarrow T_c$. For this we should start with
(\ref{Lag:no-L}) and work with broken Lorentz invariance. This means
that we have to consider the condensates like $\bar{q}\gamma_\mu
D_\nu q$ in the current correlators. It turns out however that in
HLS/VM theory such invariance breaking appears as a small correction
compared with the main term of $(1 + \frac{3(N_c^2-1)}{8N_c}\,
\frac{\alpha_s}{\pi})$ in the Lorentz-invariant matching condition
of the form
 \be &&
   \frac{F^2_\pi (\Lambda)}{{\Lambda}^2}
  = \frac{1}{8{\pi}^2} \left( \frac{N_c}{3} \right)
  \Biggl[
    1 +
    \frac{3(N_c^2-1)}{8N_c}\, \frac{\alpha_s}{\pi}
    + \frac{2\pi^2}{N_c}
      \frac{
        \left\langle
          \frac{\alpha_s}{\pi} G_{\mu\nu} G^{\mu\nu}
        \right\rangle
      }{ \Lambda^4 }
\nonumber\\
&& \qquad\qquad\qquad\qquad
    {}+ \frac{288\pi(N_c^2-1)}{N_c^3}
      \left( \frac{1}{2} + \frac{1}{3N_c} \right)
      \frac{\alpha_s \left\langle \bar{q} q \right\rangle^2}
           {\Lambda^6}
  \Biggr]\ .
 \ee
This implies that the difference between $F_{\pi,{\rm bare}}^t$ and
$F_{\pi,{\rm bare}}^s$ is small compared with their own values, or
equivalently, the bare $\pi$ velocity defined by $V_{\pi,{\rm
bare}}^2 \equiv  F_{\pi,{\rm bare}}^s / F_{\pi,{\rm bare}}^t$ is
close to one. We will give an estimate of the correction to this
bare pion velocity later.

Now given the result that $V_{\pi,{\rm bare}}=1$, we need to compute
the quantum corrections so as to compare with nature. Here Sasaki's
non-renormalization theorem~\cite{NonRen-sasaki} will help. The
argument for the theorem goes as follows. At $T\ll T_c$, the pion
velocity -- denoted $v_\pi$ for the physical quantity -- receives a
hadronic thermal correction from the pion field of the form
\begin{eqnarray}
 v_\pi^2 (T) &\simeq&
 V_\pi^2 - N_f \frac{2\pi^2}{15}\frac{T^4}{(F_\pi^t)^2 M_\rho^2}
\nonumber\\
&& \mbox{for} \quad T < T_c. \label{low T}
\end{eqnarray}
Here the longitudinal component of the $\rho$ field (called $\sigma$
in Harada-Yamawaki theory) is suppressed by the Boltzmann factor
$\exp [-M_\rho / T]$, and hence only the pion loop contributes to
the pion velocity. Now approach $T_c$. Then the vector meson mass
drops toward zero due to the VM and the Boltzmann factor $\exp
[-M_\rho / T]$ is no longer a suppression factor. Thus at tree
order, the contribution from the longitudinal vector meson
($\sigma$) exactly cancels the pion contribution. Similarly the
quantum correction generated from the pion loop is exactly canceled
by that from the $\sigma$ loop. Accordingly we conclude
\begin{equation}
 v_\pi(T) = V_{\pi,{\rm bare}}(T)
 \qquad \mbox{for}\quad T \to T_c.
\label{phys=bare}
\end{equation}
In sum, the pion velocity in the limit $T \to T_c$ is protected by
the VM against both quantum amd hadronic loop corrections at one
loop order~\cite{NonRen-sasaki}. This implies that
$(g_L,a^t,a^s,V_\pi) = (0,1,1,\mbox{any})$ forms a fixed line for
four RGEs of $g_L, a^t, a^s$ and $V_\pi$. When a point on this fixed
line is selected through the matching procedure (this is explained
in detail in \cite{HKRS:pv}), that is to say that when the value of
$V_{\pi,{\rm bare}}$ is fixed, the present result implies that the
point does not move in a subspace of the parameters. Approaching the
chiral symmetry restoration point, the physical pion velocity itself
will flow into the fixed point.

The corrections due to the breaking of Lorentz invariance to the
bare pion velocity are model-dependent and cannot be pinned down
accurately. However a rough estimate shows that they are small. With
a wide range of QCD parameters, $\Lambda_M = 0.8 - 1.1\,
\mbox{GeV}$, $\Lambda_{QCD} = 0.30 - 0.45\, \mbox{GeV}$ and the
range of critical temperature $T_c = 0.15 - 0.20\, \mbox{GeV}$, it
has been found that
\begin{equation}
 \delta_{\rm bare}(T_c) = 0.0061 - 0.29 .
\end{equation}
Thus we find the $bare$ pion velocity to be close to the speed of
light:
\begin{equation}
 V_{\pi,{\rm bare}}(T_c) = 0.83 - 0.99\,.
\end{equation}
Now thanks to the non-renormalization theorem~\cite{NonRen-sasaki},
which is applicable here as well, i.e., $v_\pi (T_c)=V_{\pi, {\rm
bare}} (T_c)$, we arrive at the physical pion velocity at chiral
restoration:
\begin{equation}
 v_{\pi}(T_c) = 0.83 - 0.99\,.\label{VMvpi}
\end{equation}

The dramatic difference in predictions for $v_\pi$ near $T_c$
between the sigma model scenario, $v_\pi\sim 0$ and the HLS/VM
scenario, $v_\pi\sim 1$ should be testable by experiments or lattice
calculations.  In fact this issue has been addressed in connection
with the STAR data on HBT~\cite{wilczek}. However the result is
inconclusive.

%-----------------------------------------
\section{Other Observables}
\subsection{Relation of Brown-Rho scaling to Harada-Yamawaki vector
manifestation}
Brown-Rho scaling \cite{BR:91} was one of the first
attempts in nuclear physics to formulate medium dependent effects
associated with the approach to chiral restoration as the scale,
either with temperature or density or with both, was increased. A
simple way to see that dynamically generated masses do scale was
introduced by Lutz et al. \cite{LKW92} through the Gell-Mann,
Oakes, Renner relation
 \be
f_\pi^2 m_\pi^2 = 2\bar m \langle\bar q q\rangle,
 \ee
where $\bar m$ is the bare quark mass. Both $\bar m$ and $m_\pi$,
which is protected against scaling to the extent that it is a
Goldstone boson, do not scale. This relation would then produce
 \be \frac{f_\pi^\star}{f_\pi} = \frac{\sqrt{\langle\bar q
q\rangle^\star}}{\sqrt{\langle\bar q q\rangle}}, \label{eqfs}
 \ee
which holds quite well for low densities\footnote{To be precise,
this relation in medium is a relation for the space component of the
pion decay constant which is different from the time component since
Lorentz invariance is broken.}. In fact, for low densities one has
the relation \cite{DL90,Cohen} \be \frac{\langle\bar q
q\rangle^\star}{\langle\bar q q\rangle} = 1 -\frac{\sigma_\pi
n}{f_\pi^2 m_\pi^2} \label{eqla} \ee where $n$ is the vector
density. Eq.~(\ref{eqla}) holds to linear approximation. At higher
densities Koch and Brown \cite{KochBrown} showed that the entropy
from reduced mass hadrons fit the entropy from LGS if one had
``Nambu scaling"
 \be \frac{m_{H}^\star}{m_H} = \frac{\langle\bar q
q\rangle^\star}{\langle\bar q q\rangle} ;
 \ee
i.e., the hadron {\it in-medium} mass scaled linearly with the quark
scalar density. This scaling seems to come out in a number of QCD
sum rule calculations, also. It holds in the Harada and Yamawaki RG
theory for high temperatures or densities approaching chiral
restoration which takes place at the fixed point where $m_V^\star$
and $g_V^\star$ go to zero.

The above applies to what we call the parametric scaling; i.e., to
the scaling of the parameters $F_\pi$, etc. which enter into the
chiral Lagrangian. One must then take this Lagrangian and calculate
thermal or dense loops, which will somewhat change the medium
dependence. A point which is generally unappreciated in the
heavy-ion theory community is that in a heat bath even at low
temperatures the (second) loop corrections are mandatory for
consistency with the symmetry of QCD. In fact, in the combination of
parametric and loop terms, the pole mass of the vector meson
increases proportional to $T^4$ near zero temperature with no $T^2$
term present as required by the low-energy theorem \cite{Dey90}. As
the temperature of chiral restoration $T_{\chi SR}$ is approached,
both the bare mass term and the loop corrections go to zero as
$\langle\bar q q\rangle \rightarrow 0$. In this case the pole mass
does directly reflect on chiral structure as does Brown-Rho scaling.
{\it Only in the vicinity of $T_c$ does BR scaling manifest itself
transparently in the pole mass of the vector meson in a heat bath.}

Evaluation of $g_\pi^\star/f_\pi$ with Eqs.~(\ref{eqfs}) \&
(\ref{eqla}) gives a 20\% drop in this quantity by nuclear matter
density $n_0$. This agrees with the value extracted at tree order
from pionic atoms \cite{Suzuki}. The same decrease is implied by
Brown-Rho scaling for $m_\rho^\star$. However, the dense loop
enters also here and, although small, will increase the mass a few
MeV. Thus, the decrease of $\sim 15\%$ in $m_\rho^\star$ by
nuclear matter density seems reasonable.

Harada and Yamawaki find that $m_\rho^\star$ scales linearly with
$\langle\bar q q\rangle$ as $m_\rho^\star\rightarrow 0$ at the
fixed point of chiral symmetry restoration. In fact, although
the comparison with lattice results on the entropy is relatively
crude in Koch and Brown \cite{KochBrown}, it is seen that with
temperature the scaling of the masses may begin less rapidly than
the scaling with $\langle\bar q q\rangle^\star$, but that it
quickly becomes as rapid. Brown and Rho~\cite{BR2004} found that
up to nuclear matter density $n_0$, $g$ did not
scale, but slightly above $n_0$ the ratio $g^\star/m_\rho^\star$ was
roughly constant. The ratio is constant going toward the fixed point
of Harada and Yamawaki. Thus we believe that the decrease of
$m_\rho^\star$ as $\sqrt{\langle\bar q q\rangle^\star}$ goes only
up to $n\sim n_0$ and that it then scales linearly with $\langle
\bar q q\rangle^\star$. If it decreases $\sim 20\%$ in going from
$n=0$ to $n_0$, it will then increase $\sim 2\sqrt{2}$ in going
from $n_0$ to $2 n_0$, and $m_\rho^\star$ will go to zero at
$n\sim 4 n_0$, the scalar density at chiral restoration. From this
estimate we believe \be n_{\chi SR} \sim 4 n_0. \ee

Given the Walecka mean field theory \cite{Walecka} and the study
of the density and temperature dependence of a system of
constituent quarks in the Nambu-Jona-Lasinio theory \cite{BMZ:87},
Brown-Rho scaling appeared quite natural, at least the scaling
with density, even a long time before its acceptance (it is not
universally accepted even now, although it has come to life
rather quickly after each of its many reported deaths).

The Walecka theory showed that the nucleon effective mass decreased
with density. Perhaps most convincing of the arguments in its favor
was that the spin-orbit term, which depends on $(m_N^\star)^{-2}$
was increased enough to fit experiment. The usual nonrelativistic
theories were typically a factor of 2 too low in spin-orbit
interaction at that time. What could be more natural than as a nucleon
dissolves into its constituents, the masses $m_Q^\star$ of the
constituent quarks decreases at the same rate as the nucleon mass
$m_N^\star$ in Walecka theory? In fact, this is what happens in
the Harada-Yamawaki theory, although it does not contain nucleons
(the effect of fermions was studied in \cite{kimetalVM} by
introducing constituent quarks). Once the density is high enough so
that constituent quarks become the relevant variables, we should go
over to a quark description, as described above and as Bernard et
al.\ \cite{BMZ:87} did. Then the constituent quark mass will
change with increasing density, going to zero the way the constituent
quark went over to a current quark as the temperature increased from
$T=125$ MeV to 175 MeV ($T_c$(unquenched)). Since at zero density
nucleons are the relevant variables, it will take some time in
adding nucleons in positive energy states before these cancel enough
of the condensate of nucleons in negative energy states so that they
can go over into loosely bound constituent quarks.

In fact, with a cut off of $\Lambda=700$ MeV, close to what we use, Bernard
et al. \cite{BMZ:87} found that in the chiral limit the quark mass went to
zero at $2 n_0$. In our scenario of chiral restoration at $n\sim 4 n_0$
outlined earlier, this then means that $n$ has to increase from 0 to
$\sim 2 n_0$ before the nucleons dissolve into constituent quarks.

\subsection{Landau Fermi-liquid fixed point and Brown-Rho
scaling}\label{FLF-BR} The meaning of Brown-Rho scaling has often
been misinterpreted in the literature for processes probing
densities in the vicinity of nuclear matter density, most recently
in connection with the NA60 dilepton data. We wish to clarify the
situation by emphasizing the intricacy involved in what the scaling
relation represents in the strong interactions that take place in
many-nucleon systems. This aspect has been discussed in several
previous publications by two of the authors (GEB and MR), but it is
perhaps not superfluous to do so once more in view of certain recent
developments. What we would like to discuss here is the connection
between the Brown-Rho scaling factor $\Phi (n)$ (to be defined
below) and the Landau parameter $F_1$ which figures in quasiparticle
interactions in Fermi liquid theory of nuclear matter. This
discussion illustrates clearly that Brown-Rho scaling cannot
$simply$ be taken to be {\it only} the mass scaling as a function of
density and/or temperature as is often done in the field. What this
illustrates is that the $\Phi$, related in an intricate way to a
quasiparticle interaction parameter in Landau Fermi-liquid theory of
nuclear matter, incorporates not just the the ``intrinsic density
dependence" (IDD in short) associated with Wilsonian matching to
QCD, a crucial element of HLS/VM, but also some of what is
conventionally considered as many-body interactions near the Fermi
surface associated with the Fermi liquid fixed point. It clearly
shows that it is dangerous to naively or blindly apply Brown-Rho
scaling to such heavy-ion processes as low-mass dileptons where the
density probed is not much higher than nuclear matter density, as was
done by several workers in QM2005. We will present arguments more specific
to dilepton processes in subsection \ref{NA60}, but what we mean will
already be clear at the end of this subsection.

\subsubsection{Chiral Fermi liquid field theory (CFLFT)}
It was argued by Brown and Rho in \cite{BR-PR-DD} (where previous
references are given) that addressing nuclear matter from the point
of view of effective field theory involves ``double decimation" in
the renormalization-group sense. The first involves going from a
chiral scale or the matching scale $\Lambda_M$ with ``bare
Larangian" to the Fermi surface scale $\Lambda_{\rm FS}$ (which will
be identified later with $\Lambda_{low-k}$). We shall describe below
a recent work~\cite{holtetal} that arrives at this result
microscopically where the $\Lambda_{\rm FS}$ will be identified with
$\Lambda_{low-k}$. How this could be achieved was discussed in a
general context by Lynn some years ago~\cite{Lynn} who made the
conjecture that the Fermi surface could arise from effective field
theories as chiral liquid soliton. For the moment, we will simply
assume that such a chiral liquid can be obtained. To proceed from
there, we exploit three observations (or, perhaps more
appropriately, conjectures). First we learn from the work of
Shankar~\cite{shankar} that given an effective Lagrangian built
around the Fermi surface, decimating fluctuations toward the Fermi
surface leads to the ``Fermi liquid fixed point" with the
quasiparticle mass $m^\star$ and quasiparticle interactions $F$
being the fixed point parameters. We next learn from Matsui's
argument~\cite{matsui} that Walecka mean field theory is
$equivalent$ to Landau Fermi liquid theory. The third observation is
that Walecka mean-field theory can be obtained in the mean field of
an effective chiral Lagrangian in which (vector and scalar) massive
degrees of freedom are present, or equivalently, an effective chiral
Lagrangian with higher-dimension operators (such as four-Fermi
operators)~\cite{gelmini-ritzi,PMR-walecka,BR-walecka}. Friman and
Rho~\cite{friman-rho96} combined the above three to write an
effective chiral Lagrangian endowed with Brown-Rho scaling that in
mean field gives {\it Landau Fermi-liquid theory at the fixed point
that is consistent with chiral symmetry}. We call this ``chiral
Fermi liquid field theory (CFLFT)" to distinguish it from the
microscopic theory of Holt et al. \cite{holtetal}.

As reviewed in \cite{BR-PR-DD}, there are two classes of effective
Lagrangians that should in principle yield the same results in the
mean field. One is closely related to a generalized HLS (GHLS)
theory where a scalar and nucleons are added to vector mesons.
Restricted to symmetric nuclear matter, it has the simple form
 \be \calL_{II} &=&
\bar{N}(i\gamma_{\mu}(\del^\mu+ig_v^\star\omega^\mu
)-M^\star+h^\star\sigma )N \nonumber\\ & &-\frac 14 F_{\mu\nu}^2
+\frac 12 (\partial_\mu \sigma)^2
+\frac{{m^\star_\omega}^2}{2}\omega^2
-\frac{{m^\star_\sigma}^2}{2}\sigma^2+\cdots\label{leff2} \ee where
the ellipsis denotes higher-dimension operators and the star refers to ``parametric density dependence" that emerges from a Wilsonian matching
to QCD of the type described by Harada and Yamawaki~\cite{HY:PR}.
We have left out (pseudo)Goldstone fields and isovector and strange vector meson fields which do not contribute at mean
field level. Note that contrary to its appearance, (\ref{leff2}) is
actually consistent with chiral symmetry since here both the
$\omega$ and $\sigma$ fields are {\it chiral singlets}. In fact, the
$\sigma$ here has nothing to do with the chiral fourth-component
scalar field of the linear sigma model except perhaps near the
chiral phase transition density where ``mended symmetry" may
intervene; it is a ``dilaton" connected with the trace anomaly of
QCD.

An alternative Lagrangian which is in a standard chiral symmetric
form involves only the pion and nucleon fields  which may be
considered as arising when the heavy mesons -- both scalar and
vector mesons -- are integrated out:
 \be
\calL_I=\bar{N}[i\gamma_{\mu}(\del^{\mu}+iv^{\mu} +g_A^\star\gamma_5
a^{\mu}) -M^\star]N -\sum_i C_i^\star (\bar{N}\Gamma_i N)^2 +\cdots
\label{leff}
 \ee
where the ellipsis stands for higher dimension and/or higher
derivative operators and the $\Gamma_i$'s are Dirac and flavor matrices
as well as derivatives consistent with chiral symmetry. Here we
reinstated the pionic vector and axial vector fields $v_\mu$ and
$a_\mu$ respectively, since the pion contributes (through exchange)
to the Landau parameters. We will go back and forth between the two
Lagrangians in our discussion.

Leaving out the details which can be found in \cite{BR-PR-DD,song},
we summarize the essential features in what is obtained for nuclear
matter. In calculating nuclear matter properties with our effective action,
the first thing to do is to determine how the nucleon and meson
masses scale near nuclear matter saturation density. This cannot be
gotten by theory, so we need empirical information. This can be done
by looking at the response of nuclear matter to external fields,
i.e., the photon. This was first done in
\cite{friman-rho96,MR:MIGDAL} using (\ref{leff}) in which the
isovector anomalous nuclear orbital gyromagnetic ratio $\delta g_l$
was expressed in terms of Brown-Rho scaling plus contributions from
the pion to the Landau parameter $F_1$\footnote{This relation is
valid up to near nuclear matter density, that is, near the
Fermi-liquid fixed point and may not be extended to much higher
densities.}
 \be
\delta g_l=\frac 49[\Phi^{-1}-1-\frac
12\tilde{F}_1^\pi], \label{gyromag}
 \ee
where $\tilde{F}_1^\pi$ is the pionic contribution to the Landau
parameter $F_1$ -- which is precisely calculable for any density
thanks to chiral symmetry -- and
 \be
\Phi (n)=\frac{m_M^\star (n)}{m_M},
 \ee
which is referred to as the ``Brown-Rho scaling factor." Here the
subscript $M$ stands for the mesons $M=\sigma, \rho, \omega$. The
isovector gyromagnetic ratio $\delta g_l$ is measured
experimentally. The most precise value comes from giant dipole
resonances in heavy nuclei~\cite{schumacher}: $\delta g_l=0.23\pm
0.03$. With $\frac 13 \tilde{F}_1^\pi=-0.153$ at nuclear matter
density $n_0$, we get from (\ref{gyromag}),
 \be
\Phi (n_0)=0.78, \label{Phi(n)}
 \ee
which is consistent with the value obtained in deeply bound pionic
atoms~\cite{yamazaki}
 \be
\frac{f_\pi^\star (n_0)}{f_\pi}\simeq 0.80.
 \ee
We should stress that this is a value appropriate for normal nuclear
matter density which should be reliable near the Fermi liquid fixed
point. For describing nuclear matter properties, we need to know how
it varies near nuclear matter equilibrium density. A convenient
parametrization is
 \be
 \Phi (n)=\frac{1}{1+yn/n_0}
\ee with $y=0.28$.

The Landau effective mass of the quasiparticle at the fixed point is
given by
 \be
m_N^\star (n)/m_N=\left(\Phi^{-1}-\frac 13\bar{F}_1^\pi\right)^{-1},
 \ee
which at the equilibrium density predicts
 \be
m_N^\star (n_0)/m_N=0.67.
 \ee
Note that the nucleon mass scales slightly faster than meson masses.
This was noted in \cite{BR:91} in terms of the scaling of $g_A$ in
medium.

We now look at other properties of nuclear matter with
(\ref{leff2}). Our construction of chiral Fermi liquid theory
instructs us to treat the Lagrangian in mean field with the mass and
coupling parameters subject to the Brown-Rho scaling. With the
standard free-space values for the $\omega$ and $\rho$ mesons and
the scalar meson mass $m_\sigma\approx 700$ MeV\footnote{Not to be
confused with the Goldstone boson $\sigma$ in HLS theory. Here it is
a chiral singlet effective field of scalar quantum number that
figures in Walecka-type mean-field theory.}, the properties of
nuclear matter come out to be~\cite{song,sbmr97}
 \be
B= 16.1\ {\rm MeV},\ \ k_F=258\ {\rm MeV},\ K=259\ {\rm
MeV},\nonumber\\ m_N^\star/m_N=0.67\label{pred1}
 \ee
where $B$ is the binding energy, $k_F$ the equilibrium Fermi
momentum and $K$ the compression modulus. The values (\ref{pred1})
should be compared with the standard ``empirical
values"\cite{furnstahl-serot}
 \be
B= 16.0\pm 0.1\ {\rm MeV},\ \ k_F=256\pm 2\ {\rm MeV},\ K=250\pm 50\
{\rm MeV},\nonumber\\
\ m_N^\star/m_N=0.61\pm 0.03.\label{emp}
 \ee
The predicted results (\ref{pred1}) are in a good agreement with
empirical values. Given the extreme simplicity of the theory, it is
rather surprising.

We should remark that what makes the theory particularly sensible is
that it is thermodynamically consistent in the sense that both
energy and momentum are conserved~\cite{song,smr-thermodynamics}.
This is a nontrivial feat. In fact, it has been a major difficulty
for nuclear matter models based on Lagragians with density-dependent
parameters to preserve the energy-momentum conservation. In the
present theory, this is achieved by incorporating a chiral invariant
form for the density operator.

\subsubsection{Brown-Rho scaling and microscopic calculation
of the Landau parameters}

In the CFLFT description given above, we relied on three
observations -- the validity of which are yet to be confirmed -- on
the connection between an effective chiral action (or an effective
chiral Lagrangian in mean field) and Landau's Fermi liquid fixed
point theory, in particular with one of the fixed point parameters
mapped to Brown-Rho scaling at the corresponding density. In a
recent work, Holt et al.~\cite{holtetal} obtained successfully a
realistic Fermi liquid description of nuclear matter in a
microscopic approach that combines the two decimations subsumed in
the CFLFT approach~\cite{sbmr97}. The approach of Holt et al. starts
with phenomenological potentials fit to scattering data up to a
momentum $\Lambda_{NN}\sim 2.1$ fm$^{-1}$.\footnote{This momentum
corresponds to the Berkeley relative momentum from the 350 MeV
laboratory energy in the cyclotron.} To understand their result, we can
recast their argument in terms of the HLS theory that we are using.
There is no such potential built from HLS Lagrangian in the
literature. However we expect, based on the work of Bogner et
al.~\cite{bogneretal}, the resulting driving potential $V_{low-k}$
to be qualitatively the same for the HLS and phenomenological models
for low-energy processes. This, we suggest, is essentially the
manifestation of the power of what is called ``more effective
effective theory" (or MEEFT for short) explained in \cite{BR-PR-DD}.

Let us imagine that we have a generalized hidden local symmetry
(GHLS) theory discussed in section \ref{GHLS} that contains a
complete set of relevant degrees of freedom for nuclear matter, say,
$\pi$, $\rho$, $\omega$, $a_1$, $\sigma$, etc., matched to QCD at a
matching scale $\Lambda_M$. There are no explicit baryon degrees of
freedom in this theory. However as discussed in section \ref{HDQCD},
baryons must emerge as skyrmions. Since no description of nuclear
dynamics starting from a GHLS exists -- and we see no reason why it
cannot be done -- one can alternatively introduce baryon fields as
matter fields and couple them in a hidden local symmetric way. This
is what one does in standard chiral perturbation theory with global
chiral symmetry with pions and nucleons as the only explicit degrees
of freedom. The ``bare" Lagrangian obtained by Wilsonian matching
will carry such parameters as masses and coupling constants endowed
with an ``intrinsic" background (temperature or density) dependence.
These are the quantities that track the properties of the quark and
gluon condensates in medium, and hence Brown-Rho scaling.

Given the ``bare" Lagrangian so determined, one can then proceed in
three steps:
\begin{enumerate}
\item First one constructs NN potentials in chiral perturbation theory
with the vector mesons treated \`a la Harada and
Yamawaki~\cite{HY:PR} -- here hidden local symmetry plays a crucial
role even at zero density as emphasized in \cite{HY:PR}. The chiral
perturbation procedure is as well formulated in HLS theory as in the
standard approach without the vector degrees of freedom.
\item Next one performs a (Wilsonian) renormalization-group
decimation to the ``low-k scale"\footnote{In this scheme, this
``low-k scale" corresponds to $\Lambda_{FS}$ introduced above.}
$\Lambda_{low-k}\approx 2\ {\rm fm}^{-1}\sim \Lambda_{NN}$ to obtain
the $V_{low-K}$. As stated, we expect the result for $V_{low-k}$ to
be basically the same as that obtained in
\cite{bogneretal,schwenketal} for the T matrix for NN scattering for
which matter density is low. However it will differ in medium due to
the intrinsic background dependence which is missing in
\cite{schwenketal}. This step will correctly implement the {\it
first decimation} of \cite{BR-PR-DD} not only in free-space but also
in dense medium. The intrinsic dependence incorporated at this stage
is missing in $all$ works found in the literature.
\item Finally one feeds the $V_{low-k}$ so determined into the B\"ackman, Brown and Niskanen nonlinear equation (their equation (5.3))~\cite{backman},
which resulted from the truncation of the Babu-Brown \cite{babu-brown}
equation in the sum over Fermi Liquid parameters to $l=0$ and 1, and solves it by iteration. (For a Green's function
formalism for Landau Fermi liquid theory, see \cite{abrikosov}.) The
Babu-Brown equation introduces the induced interaction into
dynamical calculations involving Fermi liquid theory and has its own
renormalization group treatment \cite{schwenketal} which has very
successfully been carried out for neutron matter.
\end{enumerate}
These procedures will lead to the Landau parameters given as the sum
of a ``driving term" and an ``induced term." Given the Landau
quasiparticle interactions so determined, the standard Fermi liquid
arguments are then applied to computing the energy density, Landau
effective mass, compression modulus, etc. that
describe nuclear matter. Now had the above three-step
procedure been followed with HLS, the theory would have Brown-Rho
scaling automatically incorporated. However since Holt et al.
\cite{holtetal} take, for step 1, phenomenological potentials in
which the intrinsic density dependence (IDD) required by matching to
QCD is missing, they need to implement the IDD by hand. They find
that without IDD, the known properties of nuclear matter compression
modulus, etc. cannot be reproduced correctly. To incorporate
Brown-Rho scaling into the potential they employ, e.g., the Bonn
one-boson-exchange potential, they introduce $\sigma$-tadpole
self-energy corrections to the masses of the nucleons and exchanged
bosons. Although this procedure may lack the consistency achieved
through Wilsonian matching, it should however be equivalent to
Brown-Rho scaling in its simplest form. With the Brown-Rho scaling
suitably implemented, the result obtained in \cite{holtetal} comes
out to be quite satisfactory.

There are several observations one can make from this result. 1) One
can think of this as a confirmation of the soundness of the double
decimation procedure. 2) Both the microscopic
approach~\cite{holtetal} and the effective theory
approach~\cite{sbmr97} -- which complement each other -- indicate
the importance of Brown-Rho scaling in the structure of nuclear
matter. 3) There must be a relation -- most likely quite complicated
-- between $\Phi$ and the microscopic potential $V_{low-k}$ valid at
nuclear matter density. We believe this relation to result from the
scale invariance in nuclear phenomena which results when pion
exchange is unimportant in nuclear phenomena, as reviewed in Brown
and Rho \cite{BR-PR-DD}. There are very few places in {\it nuclear
spectra} where the pion plays the main role\footnote{One should,
however, note that this is not the case in nuclear response
functions, namely in nuclear matrix elements of electroweak
currents. It is known that in certain transition matrix elements,
such as in M1 transitions and axial charge transitions, soft-pions
play an extremely important role. This is referred to as ``chiral
filter mechanism~\cite{NRZ}."}. For instance, it does not play much
of a role in the polarization phenomena reviewed by Brown and Rho
\cite{BR-PR-DD}, so they come out to be in some sense ``scale
invariant" for low densities. The pion is, of course, protected from
mass change by chiral invariance, and, therefore, does not
participate in this ``scale invariance." However, in the
second-order tensor interaction which is of primary importance for
the saturation of nuclear matter, the $\pi$ and $\sigma$ play
counterpoint; they enter incoherently, their coupling having
opposite sign. The dropping of the $\rho$-mass in Brown-Rho scaling
therefore cuts down the tensor interaction greatly, running up the
compression modulus substantially. In the usual calculations which do
not employ the intrinsic ({\it in-medium}) dependence of the masses
on density, this effect is included empirically as a three-body
interaction. This seems to be the only place in nuclear spectra
where Brown-Rho scaling seems to be really needed {\it explicitly}
in the nuclear many-body problem, although Holt et al.
\cite{holtetal} find it helpful to include Brown-Rho scaling in
terms of a scalar tadpole in order to bring the effective mass
$m_N^\star$ down towards the lower Walecka values. Nuclear physics
have, however, lived with a substantial spread in effective masses
$-$ which cannot be measured directly in nuclear physics $-$ for
years, so what we consider as an improvement in $m_N^\star$ is not
universally accepted. This reflects again on an intricate interplay
between Brown-Rho scaling and many-body interactions.

As a final remark in this subsection, let us return to the
implications of the manifestation of chiral symmetry in dense (or
hot) medium. It is clear that the connection between Brown-Rho
scaling and many-body interactions is highly intricate, particularly
near nuclear matter density, and needs to be carefully assessed case
by case. As we learned from HLS with the vector manifestation fixed
point, Brown-Rho scaling can serve as a clean-cut litmus for chiral
restoration -- a matter of intense current interest in heavy ion
physics -- only near the critical point. Only very close to the
critical point is the scaling factor $\Phi$ directly locked to the
chiral order parameter $\la\bar{q}q\ra^\star$. Far away from that
point,  particularly near normal nuclear matter density, the
connection, strongly infested with many-body interactions, can be
tenuous at best. To the extent that the dileptons in NA60 as well as
in CERES, for instance, do not selectively sample the state of
matter near the chiral transition point, information on the order
parameter cannot be extracted cleanly from the measured spectral
function.

\subsection{Fluctuating around the VM fixed point}
The traditional approach to hadronic physics in (hot and/or dense)
medium has been to start with a Lagrangian with appropriate
symmetries built on the matter-free vacuum and incorporate medium
effects mostly at low orders in perturbation theory. As stated at
several places in our work, while this can be a valid procedure when
the system is near $T=n=0$, it is not at all clear how such
low-order calculations that do not account for the intrinsic background
dependence can be trusted when the temperature and/or density is
near the critical point. Now in the framework of hidden local
symmetry theory with the VM, there is another point from which one
can do calculations accurately and systematically, namely near the
VM fixed point. The strategy of fluctuating around the VM fixed
point was discussed in some detail in the Nagoya
lecture~\cite{MR:Nagoya}. As discussed there, even certain
properties of hadrons in matter-free space that highlight chiral
symmetry -- such as for instance the chiral doubling in heavy-light
hadrons -- can be more readily treated if started from the VM. It is,
however, for processes that take place near the chiral restoration
point that fluctuating around the VM fixed point will prove to be a
lot more powerful and efficient. One specific example is the
electron-driven kaon condensation in dense stellar
matter~\cite{kaoncond}.

\subsubsection{Kaon condensation treated from the VM fixed point}
Kaon condensation treated with a theory constructed around the
$T=n=0$ vacuum, much discussed in the literature, is beset with many
complications associated with the strong coupling involved in the
interactions. Here we briefly describe how we can avoid the plethora
of complications and zero-in on the main mechanism of kaon
condensation when treated from the vector manifestation fixed point.

What is the problem in starting from the $T=n=0$ vacuum? Here one
typically does chiral perturbation theory with a chiral Lagrangian
defined for elementary interactions around the vacuum. Several
complications arise in doing so. The first is the problem of
identifying the relevant degrees of freedom and deciding
how to treat them. For instance, $\Lambda (1405)$ plays an
important role in the anti-kaon--nucleon interaction near
threshold~\cite{weiseetal} and hence should be important when
low-energy kaon-nucleon interactions are to be taken into account.
Depending upon the energy involved near the threshold, the kaon-nucleon interaction could be repulsive or attractive. This would
imply that one would have to do multichannel calculations to
correctly account for the interactions as density moves up from
zero. Higher order chiral perturbation terms can become very
important which means that one would have to face a theory in which
there can be a large number of undetermined (free) parameters, with
the attendant loss of predictive power. Various mechanisms that
cannot be accounted for in low order chiral perturbation series,
such as short-range correlations between nucleons~\cite{pandha}, can
have drastic effects on determining the critical density at which
kaons can condense. At present, there is no systematic way to assess
whether these effects survive high-order treatments consistent with
the premises of QCD. We suggested in \cite{BLPR:06} that these
mechanisms are ``irrelevant" in the RGE sense in the density regime
where kaons condense\footnote{An example is the four-Fermi
interaction that involves $\Lambda(1405)$ which we might classify as
``dangerously irrelevant" in the sense used in condensed matter
systems~\cite{senthil}: Certain high dimension perturbations are
found to be important in the paramagnet phase but irrelevant at the
critical point. In the same vein, the four-Fermi interaction
important near the $KN$ threshold is important for triggering kaon
condensation but highly irrelevant for determining the critical
density.}. Assuming that kaons condense in the vicinity of the
chiral restoration point, which is the vector manifestation fixed
point, it was suggested to be more expedient to start from the VM
fixed point to locate the kaon condensation point.

In doing this, there is a subtle point that needs to be attended to.
Seen bottom-up, once kaons condense, the state is no longer the
usual Fermi liquid making up nuclear matter. As discussed in
\cite{kimetal}, the Fermi seas of the up and down quarks get
distorted from that of a Fermi liquid with, among others, the
isospin symmetry and the parity spontaneously broken due to the
condensate $\la K\ra$ etc., and it is not clear whether and how this
state makes a transition to the Wigner mode in which chiral
symmetry is restored. This conundrum is avoided in the kaon
condensation scenario suggested in \cite{BKRT}, which we believe is
the scenario chosen by compact stars. In this scenario, electrons
with high chemical potential, not directly involved in the strong
interaction that leads to the flow towards the VM fixed point,
induce the ``crash" by decaying into kaons at a density below that
of the chiral phase transition. The RGE flow up to the crash point
is dictated by HLS dynamics and hence ``knows" about the VM fixed
point and its location in density, assuming that the electrons do not
distort the strong interactions. Were it not for the electrons, the
HLS/VM would most likely be irrelevant since the flow \`a la HLS/VM
would be stopped when kaons condense.

Seen from the HLS/VM point, the only relevant interactions between
kaons and fermions -- which would be quasiquarks instead of baryons
-- would be via exchanges of the light vector mesons, $\rho$ and
$\omega$\footnote{In HLS/VM, the $a_1$'s are joined by a scalar $s$
at the critical point. One would expect from Weinberg's mended
symmetry~\cite{weinberg} that these degrees of freedom will also
become massless in the chiral limit at the critical point as
indicated in GHLS~\cite{HS:a1,Hidakaetal}, so the question could be
raised as to how these degrees of freedom would enter in the process
of kaon condensation. We have no clear answer to this question at
the moment.}. With Brown-Rho scaling taken into account, the
prediction based on the VM fixed point at $n_{\chi SR} \sim 4n_0$ is
that kaons will condense at~\cite{BLPR:06}
 \be
n_K\sim 3n_0.
 \ee

The mass of the neutron star in J0751$+$1807 has been measured by
Nice et al. \cite{Nice05}. At 95\% confidence the mass is
$2.1_{-0.5}^{+0.4}\msun$. This causes a problem for our $n_K\sim 3
n_0$ scenario for the maximum neutron star mass of $1.5\msun$. Lee,
Brown and Park \cite{LBP06} indicated that probably some repulsion
between $K^-$-mesons in the condensate after the neutron star is born
should be introduced, in order to take into account the short-range
repulsion between the $K^-$-mesons which are made up out of fermions
$-$ the quarks $-$ and therefore should experience some van der
Waals type repulsion at short distances. Since the $K^-$ mesons are
small in extent and relatively diffuse, this repulsion would not be
expected to be large. Lee, Brown and Park \cite{LBP06} show that the
maximum neutron star mass could be raised to $1.7\msun$ without
upsetting the pattern of well-measured relativistic neutron star
binaries, but with a higher stable mass for neutron stars, most
binaries would end up with pulsar mass several tenths of a solar
mass greater than the companion mass, which is not seen in the
binaries.

\subsection{Dilepton production}\label{NA60}

In Quark Matter 2005 (proceedings to be published in Nuclear
Physics) the calculations of Ralf Rapp in what Rapp interpreted as
various different scenarios such as dropping masses, enhanced
widths due to many-body nuclear interactions, etc. were compared with
the dilepton data of the collaboration NA60, with the conclusion
that the dropping mass scenario was disproved. We claim that to conclude, therefore, that Brown-Rho scaling is also disproved is totally wrong.
As discussed above in section \ref{FLF-BR}, in the
way formulated here, Brown-Rho scaling {\it near nuclear matter
density} contains a variety of different aspects of nonperturbative
QCD dynamics consisting not only of intrinsic background dependence
characterized by the quark condensate -- the order parameter -- but
also of certain aspects of many-body dynamics associated with the
Fermi surface (in the case of dense medium). As mentioned before, the situation is vastly
clearer and more transparent in the HLS framework {\it only near} the critical
point. Thus measurements that sample mainly the vicinity of normal
nuclear matter density cannot be directly related to effects
signalling partial chiral symmetry restoration. Our assertion here
will be that dilepton measurements so far performed, including the
NA60, have not probed the regime where the signal can unambiguously
be seen. It would be premature to make a conclusion regarding the
validity of Brown-Rho scaling based on the predictions of a
naive dropping mass scenario.

The hope of the dilepton experiments of the CERES and NA60 type was
to probe the chiral symmetry structure of the vacuum as temperature
and/or density is varied. What transpires clearly from the NA60
measurement is that the spectral function in the vector meson
channel (or in fact any channel) involves a plethora of assumptions
whose validity is not fully under control, compounded with a
multitude of different reaction mechanisms that are difficult to
separate. Among others, even assuming thermal equilibrium which
justifies the notion of temperature, how the system produced at high
temperature and/or density evolves in the expansion until the
dileptons are detected must play a crucial role but is not at
present fully controlled. The measured quantities at a given
temperature, density, etc.\ contain a mixture of various parameters
that are not directly connected to the intrinsic QCD properties
that one wants to extract. This means that in order to address the
purported issue of chiral symmetry, one would have to have a
systematic and consistent theoretical tool to sort out all the
elements that enter in the analysis of the experiment. Furthermore,
one needs to incorporate the experimental conditions, such as the
cuts, etc.\ before one can confront the data with theoretical
predictions.

We do not have the necessary codes to make calculations that can be
compared with the data. But we know what theoretical ingredients
have to be included in the framework we have developed. It has been
pointed out~\cite{BR:NA60} that the minimum ingredients for testing
Brown-Rho scaling in $any$ process are three-fold: (1) the subtle
interplay between the intrinsic background dependence in the parameters
of the HLS Lagrangian, which follows from matching EFT to QCD, and
quantum loop corrections with that Lagrangian, (2) the essential
features of the vector manifestation, e.g., the violation of the
vector dominance assumption as temperature and/or density approaches
the critical point, and (3) the many-body nature of the system with
Fermi surfaces that is inherent in the HLS Lagrangian in medium, e.g.,
the ``fusing" of Brown-Rho scaling and ``sobar"
configurations~\cite{RW:00} that arise from particle-hole
excitations. All of these elements are intricately connected. {\it
Given the intricacy involved -- which we admit is not easy to
unravel fully -- our conclusion is that dilepton processes in the
presently available kinematics of heavy-ion collisions are not --
contrary to what has been believed -- a clean snapshot for chiral
properties of hadrons in medium.}

We give a brief summary of these different ingredients. The last
effect is approximately accounted for but the first two are
absent in the theoretical calculations presented at QM 2005.

\vskip 0.2cm

$\bullet$ {\it (1) Parametric background dependence}

At present, the systematic temperature dependence or density dependence
of the parameters of the Lagrangian, such as masses and gauge
coupling constants, that is required by matching HLS to QCD is not
known reliably except very near the $T=n=0$ ``vacuum" and the VM fixed
point (the same consideration holds in GHLS theory). Thus the
temperature and/or density dependence consistent with HLS/VM of the
spectral functions in the pertinent channels is not available for
analyzing reliably the NA60 data. However as mentioned above, we can
make a good guess from lattice results as to how they interpolate
between the two end points for the case of temperature and from
nuclear phenomenology for the case of density. To summarize, in
terms of the temperature, the gauge coupling $g$ that controls all
other parameters of the Lagrangian changes little between $T=0$
and the flash temperature $T_{flash}\sim 125$ MeV, at which point the soft
glue starts melting and then drops roughly linearly in the quark
condensate to zero (in the chiral limit) at $T_c\sim 175$ MeV.
With Harada-Sasaki's thermal loop
corrections~\cite{HS:loop}, {\it indispensable for satisfying
low-energy theorems}, one can estimate roughly that the $\rho$ mass
does not move much from its free-space value up to $T_{flash}$,
after which it drops to zero proportional to $g$ at $T_c$. With
density, it has been shown (see \cite{BR2004} for a review) that $g$
does not change appreciably up to normal nuclear density $n_0$ and
starts dropping linearly in quark condensate slightly above $\sim 2
n_0$ (the precise value is neither known nor very important for our
purpose).

This parametric dependence is missing in $all$ of the analyses of
dilepton data available in the literature. This is a serious defect
if one wants to study chiral symmetry in dilepton production in
heavy-ion collisions.
 \vskip 0.2cm

$\bullet$ {\it (2) Vector dominance violation}

As shown by Harada and Sasaki~\cite{HS:loop}, vector dominance can
be badly broken near the critical temperature. As mentioned in
\cite{BR:NA60}, in dense medium the parameter $a$ in HLS theory tends
to quickly approach 1\footnote{This point was discussed in a more general
setting in \cite{MR:Nagoya}.}. This means that density also breaks
vector dominance. To take this VD violation into account, we need to
determine which fixed point of the three uncovered in GHLS theory is
arrived at by temperature at chiral restoration. We will consider
the VM and Hybrid types here in which vector dominance is seriously
violated.

The net effect is that the photon coupling in the dimuon production
could be cut down by a factor as large as 4. This effect is also
missing in $all$ presently available theoretical works on dilepton
production.

\vskip 0.2cm

 $\bullet$ {\it (3) ``Sobar" configurations}

The presence of the Fermi surface in many-nucleon systems requires that
in addition to the elementary mesonic excitations of the $\rho$ (and
$\omega$) quantum numbers, there can be collective particle-hole
excitations of the same quantum numbers. This means specifically
that the spectral function in the $\rho$ channel, say, will receive
contributions from, among others, both the elementary $\rho$ meson
and the ``rhosobar" (e.g., $N^*(1520)N^{-1}$) modes. In a Lagrangian
description, such modes can be included perturbatively given a
Lagrangian that explicitly contains the relevant excited baryons.
However, constructing such a Lagrangian fully consistent with QCD
is a daunting task that has not yet been achieved. So far what has been
done is a phenomenological Lagrangian calculation, that
is, at tree order. This is the Rapp-Wambach description~\cite{RW:00}.
This may be justified in very dilute systems but is highly suspect
in dense medium since the two effects (1) and (2) mentioned above,
which we consider to be crucial for the issue of chiral symmetry
structure, are missing. As explained in \cite{HY:PR,MR:Nagoya},
without local gauge symmetry, it is not obvious how to consistently
account for them in that approach.
%The best one can do is to ``fuse" the two
%configurations as reviewed in \cite{BR2004}, with Brown-Rho scaling
%taking into account both effects in an approximate way.

The most satisfactory approach to account for this many-body aspect
would be to work directly with the HLS Lagrangian consisting only of
mesonic fields, perhaps with an infinite tower, with baryons
appearing as skyrmions. Unfortunately, as mentioned, skyrmions with
integer baryon numbers are problematic near the chiral phase
transition, since the relevant fermionic degrees of freedom are more
likely quasiquarks subject to Brown-Rho scaling. As mentioned, the
difficulty is that there seem to be no stable quasiquarks, that is,
qualitons. An alternative  -- and more workable -- approach would
be, as sketched in \cite{kimetal-sobar}, to take into account the
presence of the Fermi sea by introducing ``sobar" configurations. This
would correspond to a bosonization of particle-hole configurations
in a spirit similar to the bosonization of the Landau Fermi liquid
system~\cite{marston}.

It remains to be seen how a consistent HLS calculation taking into
account the above ingredients combined with thermal/dense loop
corrections fares with the dilepton data. Such a calculation -- and
only such a calculation -- will carry well-defined and meaningful
information on the chiral symmetry property of the matter. In the
present situation, we feel that the best one can do is to ``fuse"
the two configurations (``elementary" and ``sobar") as reviewed in
\cite{BR2004}, with Brown-Rho scaling taking into account both
effects in an approximate way.

\section{Conclusions}
In this review, an argument is presented that Harada-Yamawaki hidden
local symmetry theory emerges naturally as a truncation of an
infinite tower of hidden local fields present in the holographic
dual approach to QCD. Restricted to the lowest member of the
infinite tower and Wilsonian matched to QCD (Yang-Mills gauge theory
with fundamental quark fields), the resulting
HLS is known in the chiral limit to have the vector manifestation
fixed point to which hadronic matter is driven when chiral
restoration takes place. We have argued that at this point, $\sim$
32 light degrees of freedom become massless approaching $T_c$ from
below and account for the entropy found in lattice calculations. We
have presented several cases where this scenario came out to be
consistent with the notion of the VM, in particular in the form of
``hadronic freedom" from the critical point down to what we called the
``flash point."

This prediction hangs crucially on the vector manifestation which
holds strictly in HLS with the $\rho$ and $\pi$ and possibly the
$a_1$. We should stress that this is not just an artifact of a
theory. It should be falsifiable by lattice measurements of the
vector meson mass near the critical point.

The principal conclusion that we arrive at is that physics is
$continuous$ across the chiral transition point $T_c$ and $n_c$. We
have given the argument that we see in RHIC data the indication
that $\sim 32$ light degrees of freedom go top-down across $T_c$,
changing ``smoothly" over from Wigner-Weyl mode to Nambu-Goldstone
mode. In the modern parlance, we may call this ``quark-hadron
continuity"~\cite{sw}, but it has also been referred to as the ``Cheshire
cat principle"~\cite{NRZ,cheshirecat}. The earliest hint of this
phenomenon was in the B\'eg-Shei theorem~\cite{beg-shei} mentioned
above. How these modes wind up at $T\gg T_c$ or $n\gg n_c$ with a
``perfect liquid" or a ``color-flavor-locking" is a matter we have
not addressed in this paper.

This ``smooth" movement in the character of chiral symmetry is
clearly manifest near nuclear matter density in that what represents
Brown-Rho scaling is embedded in the Landau quasiparticle
interactions. Even in such relativistic heavy-ion processes as CERES
and NA60, there is no simple delineation of effects that signal how
chiral symmetry manifests itself from mundane nuclear many-body
effects.

There are several points we need to address to strengthen our main
thesis.

The first is whether the VM argument made at one-loop order in the
renormalization group equations survives at higher-loop order. Since there
are no higher-loop calculations -- a formidable task -- at present,
we cannot give a direct answer. However, we can make a convincing
argument why the vector manifestation (VM) fixed point $(g,a)=(0,1)$
is unaffected by higher order graphs. One powerful way to show this
is to use the protection by ``enhanced symmetry" at the fixed point,
but there is a simpler way to explain it. Although two-loop or
higher-loop calculations are not available, it has been proven by
Harada, Kugo and Yamawaki~\cite{HKY} that the tree-order low-energy
theorems remain rigorously valid to all orders. In particular, the
dimension-2 operators in the effective action remain the same to all
orders. This means that the crucial relation in HLS, i.e.,
$m^2_\rho=af^2_\pi g^2$, holds to all orders. The mass, therefore,
goes to zero as $g$ goes to zero. Now the matching condition at the
matching scale $\Lambda$ says that $g=0$ when
$\langle\bar{q}q\rangle=0$ and since $g=0$ is a fixed point of the
RGE for $g$ at any order (higher loops bring in higher powers in $g$
in the beta function), it will flow to zero at the point where the
condensate vanishes. One can also see that $a=1 + {\cal O}(g^{2n})$
near $\langle\bar{q}q\rangle=0$ where $n$ is the number of loops, so
near $T_c$ the correction to 1 is small and at $T_c$, $a=1$.
Therefore we have the VM fixed point intact. We thus conclude that
all our arguments made in the vicinity of $T_c$ where the hadronic
freedom is operative hold to all orders.

The next question is whether one cannot arrive at the VM in a QCD
sum rule approach. In fact, the QCD sum rule calculation in medium
by Hatsuda and Lee~\cite{hl} is often cited as early theoretical
evidence for a dropping vector meson mass in hot/dense matter, and
since then there have been a large number of papers written on the
subject, the most recent being in connection with the NA60 dimuon
data~\cite{rrm}. As far as we know, there is only one publication on
the subject in which the vector meson mass going to zero is
$directly$ associated with chiral restoration at $T_c$ (in the
chiral limit)~\cite{ab}. Here the vanishing of the mass is
attributed to the vanishing of the quark condensate, but it is not
clear that that is associated with the vanishing of the vector
coupling which is the origin of the VM.

Broadly speaking,  we do expect, based on the recent development in
holographic gravity-gauge duality, that one should in principle be
able to arrive at the VM via QCD sum rules. In HLS/VM, the VM is
established by equating the vector-vector correlator to the
axia-vector-axial-vector correlator -- matched between HLS and QCD
at the matching scale $\Lambda$ -- when the quark condensate
vanishes and then decimating the correlators \`a la Wilson to the
relevant scale. What enables the Harada-Yamawaki theory to do this
is the hidden local gauge invariance. Now it is plausible that
hidden local symmetry results from $emerging$ holography, in which
case the problem could be addressed in terms of QCD sum rules
exploiting the infinite tower of gauge fields for the correlators as
suggested by Earlich et al.~\cite{earlich}. This also suggests that
the holographic dual approach could unravel the structure of the
states just below and just above $T_c$ that we have discussed in
terms of HLS/VM (below) and lattice indications (above).

%%%%%%%%%%%%%%%%%%%%%%%%%%%%%%%%%%%%%%%%%%%%%%%%%%%%%%%%%%%%%%%%%
\section*{Acknowledgments}
We are very grateful to the Bielefeld lattice gauge group of
Frithjof Karsch, Dave Miller, Peter Petreczky, Olaf Kaczmarek and
Felix Zantow, not only for the generous sharing of their
SU(2)$\times$SU(2) full QCD, but also for helpful criticism and help
and suggestions as to how to parameterize and how to understand
their results. We are equally grateful to Masayasu Harada and
Chihiro Sasaki for their tuition on hidden local symmetry theory on
which the developments described in this review are based. GEB was
supported in part by the US Department of Energy under Grant No.
DE-FG02-88ER40388. CHL was supported by the Korea Research
Foundation Grant funded by the Korean Government(MOEHRD, Basic
Research Promotion Fund) (KRF-2005-070-C00034).
%-------------------------------------------------------

%%%%%%%%%%%%%%%%%%%%%%%%%%%%%%%%%%%%%%%%%%%%%%%%%%%%%%%%%%%%%%%%%%%%%%%%%%%%%%%%

%%%%%%%%%%%%%%%%%%%%%%%%%%%%%%%%%%%%%%%%%%%%%%%%%%%%%%%%%%%%%%%%%%%%%%%%%%%%%%%%

\end{document}